\RequirePackage[2024-12-01]{latexrelease}
\documentclass[showpacs,showkeys,11pt,
preprint,preprintnumbers,nofootinbib,
groupedaddress,superscriptaddress,amsmath,amssymb]{revtex4}
\usepackage{xcolor}
\usepackage{tabularx}
\usepackage{amsfonts}
\usepackage{amsmath}
\usepackage{graphicx}
\usepackage{subcaption}
\usepackage{float}
\usepackage[section]{placeins}
\usepackage[export]{adjustbox}% http://ctan.org/pkg/adjustbox
\usepackage{verbatim}
\usepackage{epsfig}
\usepackage{url}
\usepackage{multirow}
\usepackage{hhline}
\usepackage{booktabs}
\usepackage{csquotes}

\newcommand {\be}{\begin{equation}}
\newcommand {\ee}{\end{equation}}
\newcommand {\ba}{\begin{eqnarray}}
\newcommand {\ea}{\end{eqnarray}}

\usepackage{tikz}
\usepackage{tikz-feynman}
\tikzfeynmanset{compat=1.1.0}
\begin{document}
\title{Search for Vector-Like Singlet Top ($T$) Quark in a Future Muon-Proton ($\mu p$) Collider at $\sqrt{s} = 5.29, 6.48,$ and $9.16$ TeV using Advanced Machine Learning Architectures}


%%%%%%%%%%%%%%%%%%%%%%%%%%%%%%%%%%%%%%%%%%%%%%%%%%%%%%%%%%%%%%%%%%%%%%%%%%%%%%%%
\author{Haroon Sagheer}
\email{haroonsagheer663@gmail.com}
\affiliation{Riphah International University, Islamabad Pakistan}

\author{M. Tayyab Javaid}
\email{ch.tayyab119933@gmail.com}
\affiliation{Federal Urdu University of Arts, Science and Technology, Islamabad, Pakistan}

\author{Ijaz Ahmed}
\email{ijaz.ahmed@fuuast.edu.pk}
\affiliation{Federal Urdu University of Arts, Science and Technology, Islamabad, Pakistan}

\author{M. Danial Farooq}
\affiliation{Federal Urdu University of Arts, Science and Technology, Islamabad, Pakistan}

\author{Jamil Muhammad}
\email{mjamil@konkuk.ac.kr}
\affiliation{Sang-Ho College \& Department of Physics, Konkuk University, Seoul 05029, South Korea}
%%%%%%%%%%%%%%%%%%%%%%%%%%%%%%%%%%%%%%%%%%%%%%%%%%%%%%%%%%%%%%%%%%%%%%%%%%%%%%%%
\date{\today}
\newcommand{\PaperAbstractText}{%

We study the sensitivity of a future $\mu p$ collider to singly produced singlet vector-like $T$ quarks in the $T\to Wb$ decay mode across both fully hadronic ($bjj$) and leptonic ($b\ell\nu$) final states at $\sqrt{s}=5.29$, $6.48$, and $9.16~\mathrm{TeV}$. Utilizing resolved-object reconstruction and TMVA-based multivariate classifiers (BDT, BDTG, MLP, and Likelihood), we perform a fast-simulation scan over $m_T=2$--$5~\mathrm{TeV}$ and the $(g^{*},m_T)$ parameter space. To quantify improvements, we compare this multivariate selection against the closest cut-based baseline at a benchmark point of $m_T=3~\mathrm{TeV}$, $\sqrt{s}=9.16~\mathrm{TeV}$, and $\mathcal{L}=3000~\mathrm{fb}^{-1}$, incorporating a $20\%$ background-normalization uncertainty via the Asimov significance $Z_A$. At this benchmark, the hadronic channel exhibits a substantial enhancement in both $S/\sqrt{S+B}$ and $Z_A$, improving from $18.05$ and $1.44$ to $273.51$ and $42.24$, respectively. Similarly, the leptonic channel improves from $3.51$ and $0.330$ to $113.85$ and $21.51$. In the full parameter scan, the $\sqrt{s}=9.16~\mathrm{TeV}$ configuration provides the strongest sensitivity at $\mathcal{L}=3000~\mathrm{fb}^{-1}$. The hadronic channel excludes (discovers) regions with $g^{*}\in[0.20,0.50]$ ($[0.30,0.50]$) up to $m_T\approx4.0~(3.5)~\mathrm{TeV}$, while the leptonic channel remains sensitive up to $m_T\approx5.0~\mathrm{TeV}$ for $g^{*}\in[0.10,0.50]$. These gains arise from the higher-purity multivariate signal regions evaluated under a common significance prescription, and not from a change of the collider configuration. The projections assume resolved-object reconstruction and fast detector simulation, and the absolute reach should be read with these assumptions in mind.
}%

\begin{abstract}
\PaperAbstractText
\end{abstract}

\maketitle
%%%%%%%%%%%%%%%%%%%%%%%%%%%%%%%%%%%%%%%%%%%%%%%%%%%%%%%%%%%%%%%%%%%
\section{Introduction}
\label{sec:intro}
%%%%%%%%%%%%%%%%%%%%%%%%%%%%%%%%%%%%%%%%%%%%%%%%%%%%%%%%%%
The Standard Model (SM) of particle physics has achieved remarkable success in explaining a wide array of experimental results, culminating in the landmark discovery of the Higgs boson at the Large Hadron Collider (LHC) \cite{Chatrchyan:2012ufa, Aad:2012tfa}. Despite this, the SM is widely regarded as an effective theory that leaves fundamental questions unanswered, particularly regarding the gauge hierarchy problem, the origin of the flavor structure, and the nature of dark matter \cite{Hill:2002ap, Altarelli:2000fu, Barbieri:1987fn}. Vector-like quarks (VLQs) emerge as a natural and compelling extension in numerous Beyond the Standard Model (BSM) frameworks designed to address these deficiencies, including composite Higgs models, little Higgs constructions, extra-dimensional scenarios, and superstring-inspired E6 models \cite{Alves:2023roadmap, Randall:1999ee, Hewett:1988xc, Arkani-Hamed:2002ncl, DeSimone:2012fs}. Unlike chiral quarks, VLQs are defined by the fact that their left- and right-handed components transform identically under the SM $SU(2)_L \times U(1)_Y$ gauge group, allowing for gauge-invariant mass terms that are not strictly tied to the electroweak symmetry breaking scale \cite{Aguilar-Saavedra:2009kvw, Branco:1986my, Cacciapaglia:2011fx}.

In many of these theoretical setups, such as the Randall-Sundrum models \cite{Randall:1999ee} or the $SO(5)/SO(4)$ composite Higgs framework \cite{Kaplan:1983fs, Agashe:2004rs}, the top-partner $T$ quark with electric charge $+2/3$ plays a critical role in stabilizing the Higgs mass by canceling quadratic divergences stemming from top-quark loops \cite{Contino:2006qr, Matsedonskyi:2012em, Perelstein:2003wd}. The phenomenology of these partners is primarily dictated by their mixing with third-generation SM quarks via Yukawa interactions, as detailed in effective descriptions of quark mixing \cite{delAguila:2000rc, Buchkremer:2013bba}. At high-energy colliders, a heavy top partner $T$ is expected to decay into a third-generation quark and a SM boson ($Wb$, $tZ$, and $tH$), with the branching fractions for the $tZ$ and $tH$ modes becoming nearly comparable at high masses due to the Goldstone boson equivalence theorem \cite{Buchkremer:2013bba, Vignaroli:2012nf, Aguilar-Saavedra:2010ze}.

Experimental searches for VLQs are a major priority for the ATLAS and CMS collaborations.  Pair-production searches have established lower mass limits approaching 1.3--1.5 TeV for various VLQ species \cite{Atre:2011ae, Sirunyan:2019sga, Aaboud:2018pii}, while single-production searches, whose reach depends on the coupling to SM quarks, extend the sensitivity to higher masses \cite{Aad:2023vlq, CMS:2026vlqWb}.  For example, a recent CMS search for a VLQ decaying to $Wb$ in single-lepton final states excludes masses below 2.4~TeV for a coupling $\kappa_W=0.2$ \cite{CMS:2026vlqWb}. While pair production of VLQs is a largely model-independent process mediated by the strong force \cite{Abada:2019lih}, single production through electroweak interactions probes the specific mixing parameters of the model and often provides a broader discovery reach at high mass scales due to its lower kinematic threshold \cite{Atre:2009rg, Vignaroli:2012nf}. However, as the mass of the top partner increases into the multi-TeV regime, its decay products become significantly boosted, resulting in highly collimated jet topologies. These ``fat jets'' present a substantial challenge for standard jet reconstruction and traditional cut-based analysis techniques, necessitating the development of more sophisticated identification strategies \cite{Almeida:2009jet, Chekanov:2024up, Ahmed:2023up, Ahmed:2025up}.

To extend the sensitivity to heavy VLQs beyond the High-Luminosity LHC, the physics community is investigating future facilities such as CLIC \cite{Han:2024clic} and muon--proton colliders \cite{Klein:2008di}.  In this work we study single production of a vector-like $T$ quark at a future $\mu p$ collider with $\sqrt{s}_{\mu p}=5.29$, $6.48$, and $9.16$~TeV in both fully hadronic and leptonic final states.  The analysis uses TMVA-based multivariate classifiers because conventional cut-based selections do not fully exploit the correlated kinematic structure of these final states.

The closest existing $\mu p$ studies of the same $T\to Wb$ topology are Refs.~\cite{Han:2025WbMuP} and \cite{Hussain:2026VLQMuP}.  Ref.~\cite{Han:2025WbMuP} provides the nearest published cut-based analysis at the same collider energies and in the same final states, using a boosted-$W$ fat-jet strategy in the hadronic channel, while Ref.~\cite{Hussain:2026VLQMuP} treats the same setup with a cut-based selection and reports representative BDT/MLP results at selected benchmark points.  The collider energies and final states considered here are therefore not new by themselves.  The contributions of the present work are:
(i)~a systematic comparison of four TMVA classifiers (BDT, BDTG, MLP, and Likelihood) in both final states, supported by overtraining checks and ROC/AUC comparisons;
(ii)~a full scan over $m_T=2$--$5$~TeV and the $(g^{*},m_T)$ plane at all three collider energies, rather than a small set of benchmark points;
(iii)~resolved-object reconstruction of the heavy-quark mass in both channels; and
(iv)~a common significance treatment, including a $20\%$ background-normalization uncertainty through the Asimov significance $Z_A$, with a like-for-like benchmark comparison to the cut-based baseline at $m_T=3~\mathrm{TeV}$, $\sqrt{s}=9.16~\mathrm{TeV}$, and $\mathcal{L}=3000~\mathrm{fb}^{-1}$.

The remainder of the paper is organized as follows.  The theoretical setup is summarized in Sec.~\ref{sec:vlq_model}.  A benchmark comparison with the closest existing cut-based studies is given in Sec.~\ref{sec:cutbased_baseline}, while the event selections, kinematic distributions, and multivariate-analysis strategy are presented in Secs.~\ref{sec:analysis_strategy} and \ref{sec:mva}.

%%%%%%%%%%%%%%%%%%%%%%%%%%%%%%%%%%%%%%%%%%%%%%%%%%%%%%%%%%%%%%%%%
\section{A Brief Overview of Vector-Like Singlet Top Quark Model}
\label{sec:vlq_model}
\subsection{Field Content and Mixing}
%%%%%%%%%%%%%%%%%%%%%%%%%%%%%%%%%%%%%%%%%%%%%%%%%%%%%%%%%%%%%%%%%
Vector-like quarks can be arranged as SU$(2)_L$ singlets, doublets, or triplets. In the
singlet case relevant for this study, the top partner $T$ carries electric charge $+2/3$ and can
mix with SM up-type quarks through Yukawa couplings to the Higgs doublet \cite{Alves:2023roadmap}. Because VLQ
mass terms are gauge invariant, a bare mass $M_T$ is allowed, and electroweak symmetry
breaking induces additional mixing between $T$ and the SM sector \cite{Buchkremer:2013bba}. This
mixing opens up the electroweak decay modes that dominate the collider phenomenology.

\subsection{Effective Interactions}
Following a standard effective parametrization for singlet VLQs \cite{delAguila:2000rc,Buchkremer:2013bba},
the interactions of a single $T$ quark with electroweak gauge bosons can be written in a
compact, model-independent form as
\be
\mathcal{L}_{T} =
\frac{g}{\sqrt{2}} \kappa_W V_{4i}\,\bar{T}_{L/R} \gamma^\mu W^+_\mu\, d_{i\,L/R}
 + \frac{g}{2 c_W} \kappa_Z V_{4i}\,\bar{T}_{L/R} \gamma^\mu Z_\mu\, u_{i\,L/R}
- \frac{\kappa_H V_{4i} M_T}{v}\,\bar{T}_{R/L} H\, u_{i\,L/R}
 + \mathrm{h.c.}
\label{eq:vlq_lagrangian}
\ee
Here $V_{4i}$ denotes the mixing between the heavy state and the SM generation $i$,
$\kappa_{W,Z,H}$ encode the coupling strengths to the $W$, $Z$, and Higgs bosons, $v$ is the
electroweak vacuum expectation value, and $c_W = \cos\theta_W$. This parametrization is
consistent with the standard VLQ framework and makes it straightforward to relate collider
observables to the underlying mixing pattern.
%%%%%%%%%%%%%%%%%%%%%%%%%%%%%%%%%%%%%%%%%%%%%%%%
\subsection{Branching Ratios}
%%%%%%%%%%%%%%%%%%%%%%%%%%%%%%%%%%%%%%%%%%%%%%%%
For the decay $T \to V q_i$ with $V = W, Z, H$, the branching fractions can be expressed as
follows ~\cite{Buchkremer:2013bba}:
\be
\mathrm{BR}(T \to V q_i) =
\frac{\kappa_V^2 |V_{4i}|^2 \Gamma_V^0}{
\sum_{j=1}^3 |V_{4j}|^2 \sum_{V' = W,Z,H} \kappa_{V'}^2 \Gamma_{V'}^0 } \, ,
\label{eq:vlq_br}
\ee
where $\Gamma_V^0$ denotes the partial width in the limit of unit coupling. In practice, the
$T \to W b$ channel is dominant for the benchmark scenarios considered here, while the $tZ$
and $tH$ modes are smaller but nearly equal at high mass, in line with the Goldstone
equivalence expectation.

%%%%%%%%%%%%%%%%%%%%%%%%%%%%%%%%%%%%%%%%%%%%%%%%
\section{Comparison with Existing Cut-Based Studies}

{

\label{sec:cutbased_baseline}

This section compares the present analysis to the closest available cut-based studies of the same $\mu p\to \nu_\mu T\bar b$ process with $T\to Wb$.  The comparison is organized so that three ingredients are kept conceptually separate: (i) the loose object-level preselection used here to define a common fiducial sample for classifier training, (ii) the optimized rectangular-cut strategies quoted in Refs.~\cite{Han:2025WbMuP,Hussain:2026VLQMuP}, and (iii) the final multivariate working point used in the present study.  The preselection is not intended as a stand-alone discovery analysis; it is only the stage at which basic detector-level acceptance and object requirements are imposed before the multivariate training.

For an explicit numerical comparison we use the benchmark point $m_T=3~\mathrm{TeV}$, $\sqrt{s}=9.16~\mathrm{TeV}$, and $\mathcal{L}=3000~\mathrm{fb}^{-1}$, because Ref.~\cite{Hussain:2026VLQMuP} quotes results at this same point and therefore allows the closest like-for-like comparison.  Ref.~\cite{Han:2025WbMuP} establishes the same topology and collider energies as the published cut-based baseline, while Ref.~\cite{Hussain:2026VLQMuP} is used below for the explicit benchmark-level numbers.  In all cases the conventional significance is quoted as $S/\sqrt{S+B}$, and the systematic robustness is evaluated with the same profile-likelihood Asimov significance of Eq.~(\ref{eq:asimov}) using a fractional background uncertainty of $\epsilon_B=20\%$.

With this setup, the comparison is tied to the same final state, collider energy, luminosity, and significance prescription as closely as the published inputs allow.  Any improvement can then be attributed primarily to the multivariate use of correlated kinematic information, rather than to a change of collider scenario or to a different uncertainty definition.

Ref.~\cite{Han:2025WbMuP} also provides an important published reach comparison at $\sqrt{s}=9.16~\mathrm{TeV}$ and $\mathcal{L}=100~\mathrm{fb}^{-1}$, where the hadronic analysis uses a boosted-$W$ strategy with a fat jet and reports a $5\sigma$ discovery reach up to about $m_T\simeq3.75~\mathrm{TeV}$ together with a 95\% CL exclusion reach up to about $m_T\simeq4.5~\mathrm{TeV}$.  That result should not be treated as a bin-by-bin surrogate for the present hadronic study, because the current manuscript reconstructs $T\to Wb\to jjb$ in a resolved small-$R$ topology and therefore absorbs high-mass merging losses differently.  For this reason the exact benchmark-level numerical comparison is anchored to Ref.~\cite{Hussain:2026VLQMuP}, while Ref.~\cite{Han:2025WbMuP} is retained explicitly as the closest published boosted-object reference for the same $\mu p$ topology.

At the published $\sqrt{s}=9.16~\mathrm{TeV}$, $\mathcal{L}=100~\mathrm{fb}^{-1}$ benchmark, the comparison with Ref.~\cite{Han:2025WbMuP} can also be stated more explicitly.  In that boosted-hadronic study, the quoted discovery and exclusion reaches are approximately $m_T\simeq3.75$~TeV and $m_T\simeq4.5$~TeV, respectively.  In the present resolved-object MVA study, the luminosity-consistent hadronic entries in Table~\ref{tab:best_vs_best_100fb} at the same collider energy give $Z=30.20$ with $Z_A=24.39$ at $m_T=4.0$~TeV, $Z=20.28$ with $Z_A=13.22$ at $m_T=4.5$~TeV, and $Z=12.44$ with $Z_A=7.85$ at $m_T=5.0$~TeV.  Under the assumptions of the present fast-simulation setup, these numbers show that the resolved multivariate strategy remains numerically competitive with the published boosted-object reach through the $4.0$--$4.5$~TeV region, while the comparison should still be interpreted with care because the two analyses reconstruct the hadronic $W$ in different ways.

\begin{table}[H]
  \centering
  
  \caption{Compact comparison of the present study with the closest existing $\mu p\to \nu_\mu T\bar b$ analyses in the $T\to Wb$ topology.}
  \label{tab:literature_comparison_summary}
  \scriptsize
  \begin{tabular}{|c|c|c|c|c|}
  \hline
  \textbf{Reference} &
  \textbf{Final states} &
  \textbf{Hadronic strategy} &
  \textbf{Quoted scope} &
  \textbf{Role in the present comparison} \\
  \hline

  \parbox[t]{2.2cm}{Ref.~\cite{Han:2025WbMuP}} &
  \parbox[t]{1.8cm}{Hadronic and leptonic} &
  \parbox[t]{2.6cm}{Boosted $W$ reconstructed as a fat jet in the hadronic channel} &
  \parbox[t]{3.0cm}{$\sqrt{s}=5.29$, $6.48$, $9.16~\mathrm{TeV}$ at $\mathcal{L}=100~\mathrm{fb}^{-1}$; at $9.16~\mathrm{TeV}$ the hadronic study reports $5\sigma$ discovery up to
  $m_T\simeq3.75~\mathrm{TeV}$ and 95\% CL exclusion up to $m_T\simeq4.5~\mathrm{TeV}$} &
  \parbox[t]{4.2cm}{Closest published reach baseline for the same $\mu p$ topology; used here to position the present resolved-object study with respect to a boosted-object hadronic analysis} \\
  \hline

  \parbox[t]{2.2cm}{Ref.~\cite{Hussain:2026VLQMuP}} &
  \parbox[t]{1.8cm}{Hadronic and leptonic} &
  \parbox[t]{2.6cm}{Conventional cut-based treatment, with representative BDT/MLP numbers at selected benchmarks} &
  \parbox[t]{3.0cm}{Explicit benchmark numbers quoted at $m_T=3~\mathrm{TeV}$, $\sqrt{s}=9.16~\mathrm{TeV}$, and $\mathcal{L}=3000~\mathrm{fb}^{-1}$} &
  \parbox[t]{4.2cm}{Closest common-benchmark numerical baseline; used for the one-point like-for-like comparison in Tables~\ref{tab:cutbased_hadronic_comparison} and
  \ref{tab:cutbased_leptonic_comparison}} \\
  \hline

  \parbox[t]{2.2cm}{Present work} &
  \parbox[t]{1.8cm}{Hadronic and leptonic} &
  \parbox[t]{2.6cm}{Resolved small-$R$ reconstruction in the hadronic channel, combined with TMVA classifiers} &
  \parbox[t]{3.0cm}{Scan over $m_T=2$--$5~\mathrm{TeV}$ and the $(g^{*},m_T)$ plane at $\sqrt{s}=5.29$, $6.48$, $9.16~\mathrm{TeV}$} &
  \parbox[t]{4.2cm}{Added value of the present study: a systematic multivariate resolved-object analysis with explicit benchmark-level and full-scan sensitivity projections under a common $Z_A$
  treatment} \\
  \hline

  \end{tabular}
  \end{table}

\subsection{Hadronic channel}
\label{subsec:cutbased_hadronic}

In the fully hadronic topology, $T\to Wb\to jjb$, the large $W\to jj$ branching fraction provides a high raw signal rate, but the analysis must suppress sizeable $\nu jj$, $\nu Wj$, $\nu Zj$, and single-top backgrounds.  At the common benchmark point, the published event yields in Ref.~\cite{Hussain:2026VLQMuP}, $S=1355.6$ and $B=4282.5$, give $S/\sqrt{S+B}=18.054$ when recalculated directly.  The corresponding signal-to-background ratio is $S/B=0.31654$, and the same $20\%$ background-normalization uncertainty used in the present paper gives $Z_A=1.437$.  This benchmark is useful because it shows that a statistically visible excess can still be fragile once residual background systematics are included.

For the same benchmark point, the present hadronic MLP working point gives $S/B=39.63$ and $S/\sqrt{S+B}=273.51$.  Using the corresponding post-selection yields in Eq.~(\ref{eq:asimov}) gives $Z_A=42.236$ for $\epsilon_B=20\%$.  Within the assumptions of the present fast-simulation study, the gain therefore comes mainly from the large increase in purity, obtained by combining information from $M_{bjj}$, $H_T$, the leading and subleading jet kinematics, $b$-jet observables, and angular separations in a correlated way.

\begin{table}[H]
\centering
\caption{{Hadronic-channel comparison at the common benchmark point $m_T=3~\mathrm{TeV}$, $\sqrt{s}=9.16~\mathrm{TeV}$, and $\mathcal{L}=3000~\mathrm{fb}^{-1}$.  The reference cut-based row is taken from Ref.~\cite{Hussain:2026VLQMuP}; the $Z_A$ values are evaluated with Eq.~(\ref{eq:asimov}) using $\epsilon_B=20\%$.}}
\label{tab:cutbased_hadronic_comparison}
\small
\begin{adjustbox}{max width=\textwidth}
\begin{tabular}{|l|c|c|c|c|c|c|}
\toprule \hline
\textbf{Analysis strategy} &
\textbf{$S$} &
\textbf{$B$} &
\textbf{$S/B$} &
\textbf{$S/\sqrt{B}$} &
\textbf{$S/\sqrt{S+B}$} &
\textbf{$Z_A(20\%)$} \\
\midrule \hline
Optimized cut-based baseline, Ref.~\cite{Hussain:2026VLQMuP} &
$1355.6$ &
$4282.5$ &
$0.31654$ &
$20.71490$ &
$18.05366$ &
$1.43724$ \\ \hline
Present MVA, MLP working point &
$76695.37$ &
$1935.29$ &
$39.63$ &
$1743.40$ &
$273.51$ &
$42.236$ \\
\bottomrule \hline
\end{tabular}
\end{adjustbox}
\end{table}

\subsection{Leptonic channel}
\label{subsec:cutbased_leptonic}

The leptonic topology, $T\to Wb\to \ell\nu b$, has a smaller branching fraction but a cleaner experimental signature.  The isolated charged lepton, sizeable missing transverse momentum, and heavy-flavour jet requirement strongly reduce the multijet contamination.  At the same benchmark point, the published event yields in Ref.~\cite{Hussain:2026VLQMuP}, $S=194.8$ and $B=2879.4$, give $S/\sqrt{S+B}=3.513$ and $S/B=0.067653$ when recalculated directly.  After including the same $20\%$ background uncertainty, the corresponding Asimov significance is $Z_A=0.3295$.  This indicates that, in a conventional cut-based treatment, the residual background remains large compared with the signal.

At the same benchmark point, the present leptonic MLP working point reaches $S/B=11.93$ and $S/\sqrt{S+B}=113.85$, corresponding to $Z_A=21.512$ for $\epsilon_B=20\%$.  Here the improvement comes from the combined use of the reconstructed $M_{b\ell\nu}$ spectrum, transverse-mass information, $E_T^{\rm miss}$, lepton kinematics, $b$-jet observables, and $\Delta R(\ell,b)$, which together define a much cleaner signal region than the cut-based baseline.

\begin{table}[H]
\centering
\caption{{Leptonic-channel comparison at the common benchmark point $m_T=3~\mathrm{TeV}$, $\sqrt{s}=9.16~\mathrm{TeV}$, and $\mathcal{L}=3000~\mathrm{fb}^{-1}$.  The reference cut-based row is taken from Ref.~\cite{Hussain:2026VLQMuP}; the $Z_A$ values are evaluated with Eq.~(\ref{eq:asimov}) using $\epsilon_B=20\%$.}}
\label{tab:cutbased_leptonic_comparison}
\small
\begin{adjustbox}{max width=\textwidth}
\begin{tabular}{|l|c|c|c|c|c|c|}
\toprule \hline
\textbf{ Analysis strategy} &
\textbf{$S$}  &
\textbf{$B$} &
\textbf{$S/B$} &
\textbf{$S/\sqrt{B}$} &
\textbf{$S/\sqrt{S+B}$} &
\textbf{$Z_A(20\%)$} \\
\midrule \hline
Optimized cut-based baseline, Ref.~\cite{Hussain:2026VLQMuP} &
$194.8$ &
$2879.4$ &
$0.067653$ &
$3.63026$ &
$3.51336$ &
$0.32953$ \\  \hline
Present MVA, MLP working point &
$14048.31$ &
$1177.56$ &
$11.93$ &
$409.39$ &
$113.85$ &
$21.512$ \\
\bottomrule \hline
\end{tabular}
\end{adjustbox}
\end{table}

The purpose of Tables~\ref{tab:cutbased_hadronic_comparison} and \ref{tab:cutbased_leptonic_comparison} is not to reinterpret the literature under new detector assumptions, but to isolate the effect of the signal-extraction strategy at a common benchmark point.  In that restricted sense, the comparison clarifies the novelty of the present study: the collider setup and final state are not new, whereas the systematic multivariate treatment over the full scan leads to markedly higher post-selection purity and hence to a more stable significance once background-normalization uncertainties are included.  For the cut-based rows, every derived quantity is recalculated directly from the published $S$ and $B$.  For the MVA rows, the displayed yields are algebraically reconstructed from the reported $S/B$ and $S/\sqrt{S+B}$ and reproduce the quoted significance measures to the displayed precision.

}

%%%%%%%%%%%%%%%%%%%%%%%%%%%%%%%%%%%%%%%%%%%%%%%%
\section{Event Selections and Kinematics Distributions}
\label{sec:analysis_strategy}
%%%%%%%%%%%%%%%%%%%%%%%%%%%%%%%%%%%%%%%%%%%%%%%%
\subsection{Hadronic Section}
%%%%%%%%%%%%%%%%%%%%%%%%%%%%%%%%%%%%%%%%%%%%%%%%
The search for a Vector-Like Top (VLT) quark singlet ($T$) in the fully hadronic channel at a muon-proton collider presents a complex experimental environment. The signal process, $\mu^- p \to \nu_\mu T \bar{b} \to \nu_\mu (Wb)\bar{b} \to \nu_\mu (jjb)\bar{b}$, results in a final state composed of at least four hard partons and significant missing transverse energy ($\mathbb{E}_T^{miss}$) from the recoil neutrino. The dominant Standard Model backgrounds are:
\begin{itemize}
    
    \item SM Single Top: $\mu^- p \to \nu_\mu t \bar{b} \to \nu_\mu (W^+ b) \bar{b} \to \nu_\mu (j j b) \bar{b}$
    \item $W$ + jets: $\mu^- p \to \nu_\mu W^+ j \to \nu_\mu (j j) j$
    \item $Z$ + jets: $\mu^- p \to \nu_\mu Z j \to \nu_\mu (j j) j$
    \item Multijet: $\mu^- p \to \nu_\mu j j$
\end{itemize}
%%%%%%%%%%%%%%%%%%%%%%%%%%%Feynmann Diagaram for Hadronic Signal%%%%%%%%%
\begin{figure}[H]
\centering
\begin{tikzpicture}
\begin{feynman}

\vertex (mu) {\(\mu^-\)};
\vertex[below=1.8cm of mu] (p) {\(p\)};

\vertex[right=2.0cm of mu] (v1);
\vertex[right=2.0cm of p] (v2);
\vertex[right=2.2cm of v1] (nu) {\(\nu_\mu\)};
\vertex[right=2.2cm of v2] (T) {\(T\)};

\vertex[right=1.4cm of T] (v3);
\vertex[above right=0.9cm and 0.9cm of v3] (j1) {\(j\)};
\vertex[below right=0.9cm and 0.9cm of v3] (j2) {\(j\)};
\vertex[below=1.0cm of T] (b) {\(b\)};

\diagram*{
    (mu) -- [fermion] (v1) -- [fermion] (nu),
    (p) -- [fermion, edge label=\(b\)] (v2),
    (v1) -- [boson, edge label=\(W^-\)] (v2),
    (v2) -- [fermion, edge label=\(T\)] (T),
    (T) -- [fermion] (b),
    (T) -- [boson, edge label'=\(W^+\)] (v3),
    (v3) -- [fermion] (j1),
    (v3) -- [fermion] (j2)
};

\end{feynman}
\end{tikzpicture}
\caption{Representative leading-order (LO) Feynman diagram for the single production of a vector-like singlet top quark in the hadronic channel at a $\mu p$ collider. The diagram highlights the underlying sub-process $\mu^- b \to \nu_\mu T$ (where the $b$ quark arises from $g \to b\bar{b}$ splitting inside the proton), followed by $T \to W^+ b$ and $W^+ \to jj$.}
\label{fig:feyn_hadronic}
\end{figure}
%\caption{Representative leading-order (LO) Feynman diagram for the single production of a vector-like singlet top quark in the hadronic channel at a $\mu p$ collider: $\mu^- p \to \nu_\mu\, T\, b$, followed by $T \to W^+ b$ and $W^+ \to j j$.}
%%%%%%%%%%%%%%%%%%%%%%%%%%%%%%%%%%%%%%%%%%%%%%%%%%%%
The dataset is simulated using MadGraph5 \cite{madgraph}, followed by PYTHIA 8.3 \cite{Bierlich:2022pythia} for parton showering and hadronization, and DELPHES 3.5 \cite{delphes} for detector simulation. To isolate this multi-TeV resonance from Standard Model (SM) backgrounds, we employ a sequential selection strategy based on nine specific kinematic observables. The motivation for the selection criteria is summarized below:
\begin{itemize} 
 \item \textbf {Object Acceptance and Trigger Thresholds:} 
 Initial event selection ensures that all physics objects fall within the detector's optimal tracking and calorimetric volume. We impose a global geometric constraint on all reconstructed jets, requiring $|\eta| < 2.5$. This reduces background contamination while preserving the signal, as reflected in the evolution of the signal efficiencies at $\sqrt{s}=5.29$, $6.48$, and $9.16$ TeV. To trigger on the hard scattering expected from a heavy resonance, we define stringent thresholds for the leading and sub-leading jets:

\item \textbf{Leading Jet Kinematics ($p_T^{j_1}, \eta_{j_1}$):} We require $p_T^{j_1} > 120$ GeV. This threshold is critical for triggering on the energetic decay products of the VLT.
\item \textbf{Sub-leading Jet Kinematics ($p_T^{j_2}, \eta_{j_2}$):} A secondary threshold of $p_T^{j_2} > 50$ GeV is applied. For high-mass signals, these two leading jets typically originate from the boosted $W$ or $T$ decay chain.
\end{itemize}
%table%%%%%%
 \begin{table}[H]
 \label{tab:I}
  \caption{Cumulative signal efficiencies for the hadronic channel across all benchmark masses $m_T=2000$--$5000$ GeV at $\sqrt{s}=5.29$ TeV.}
  \centering
  \small
  \resizebox{\textwidth}{!}{%
  \begin{tabular}{|l|r|r|r|r|r|r|r|}
  \hline
  \textbf{Selection Cuts} & \textbf{2000 GeV} & \textbf{2500 GeV} & \textbf{3000 GeV} & \textbf{3500 GeV} & \textbf{4000 GeV} & \textbf{4500 GeV} & \textbf{5000 GeV} \\
  
  \hline
  $|\eta_j| < 2.5$ & 0.50068 & 0.49836 & 0.49784 & 0.50288 & 0.49940 & 0.50016 & 0.49928 \\
  \hline
  $p_T^{j_1} > 120\,\mathrm{GeV}$ & 0.49532 & 0.49472 & 0.49508 & 0.50088 & 0.49808 & 0.49872 & 0.49828 \\
  \hline
  $p_T^{j_2} > 50\,\mathrm{GeV}$ & 0.49084 & 0.48988 & 0.48824 & 0.49092 & 0.48456 & 0.48232 & 0.47168 \\
  \hline
  $N_{\mathrm{jets}} \ge 4$ & 0.17828 & 0.16952 & 0.16064 & 0.15420 & 0.14428 & 0.13532 & 0.09748 \\
  \hline
  $N_{b\mathrm{jets}} \ge 2$ & 0.05388 & 0.04720 & 0.03932 & 0.03288 & 0.02980 & 0.02548 & 0.01600 \\
  \hline
  $H_T > 450\,\mathrm{GeV}$ & 0.05376 & 0.04712 & 0.03924 & 0.03288 & 0.02968 & 0.02548 & 0.01600 \\
  \hline
  $E_T^{\mathrm{miss}} > 10\,\mathrm{GeV}$ & 0.05316 & 0.04640 & 0.03888 & 0.03276 & 0.02948 & 0.02532 & 0.01596 \\
  \hline
  \textbf{$\epsilon$\%} & \textbf{5.316} & \textbf{4.640} & \textbf{3.888} & \textbf{3.276} & \textbf{2.948} & \textbf{2.532} & \textbf{1.596} \\
  \hline
  \end{tabular}%
  }
  \end{table}

%%%%%%%%%%%%%%%%%%%%%%%%%%%%%%%%%%%%%%%%%%%%%%%%%%%%%%%%%%%%%%%%%%%
\begin{table}[H]
\label{tab:II}
  \caption{Distribution of the Cumulative Signal Efficiencies for Hadronic Channel across all $m_{T}$ (2000-5000 GeV) points at $\sqrt{s}$ = 6.48 TeV.}
  \centering
  \small
  \resizebox{\textwidth}{!}{%
  \begin{tabular}{|l|r|r|r|r|r|r|r|}
  \hline
  \textbf{Selection Cuts} & \textbf{2000 GeV} & \textbf{2500 GeV} & \textbf{3000 GeV} & \textbf{3500 GeV} & \textbf{4000 GeV} & \textbf{4500 GeV} & \textbf{5000 GeV} \\
  \hline
  $|\eta_j| < 2.5$ & 0.49824 & 0.50124 & 0.49580 & 0.50088 & 0.49932 & 0.49820 & 0.50312 \\
  \hline
  $p_T^{j_1} > 120\,\mathrm{GeV}$ & 0.49452 & 0.49832 & 0.49380 & 0.49912 & 0.49760 & 0.49688 & 0.50220 \\
  \hline
  $p_T^{j_2} > 50\,\mathrm{GeV}$ & 0.49204 & 0.49548 & 0.49076 & 0.49480 & 0.49176 & 0.49008 & 0.49416 \\
  \hline
  $N_{\mathrm{jets}} \ge 4$ & 0.21268 & 0.20432 & 0.20016 & 0.19080 & 0.18812 & 0.18152 & 0.17756 \\
  \hline
  $N_{b\mathrm{jets}} \ge 2$ & 0.06548 & 0.05868 & 0.05248 & 0.04640 & 0.03952 & 0.03696 & 0.03236 \\
  \hline
  $H_T > 450\,\mathrm{GeV}$ & 0.06488 & 0.05860 & 0.05240 & 0.04636 & 0.03952 & 0.03696 & 0.03236 \\
  \hline
  $E_T^{\mathrm{miss}} > 10\,\mathrm{GeV}$ & 0.06412 & 0.05804 & 0.05164 & 0.04600 & 0.03916 & 0.03680 & 0.03220 \\
  \hline
  \textbf{$\epsilon$\%} & \textbf{6.412} & \textbf{5.804} & \textbf{5.164} & \textbf{4.600} & \textbf{3.916} & \textbf{3.680} & \textbf{3.220} \\
  \hline
  \end{tabular}%
  }
  \end{table}
%%%%%%%%%%%%%%%%%%%%%%%%%%%%%%%%%%%%%%%%%%%%%%%%%%%%%%%%%%%%%%%%%%%
To distinguish the massive signal from soft SM backgrounds, discriminating variables such as global energy and multiplicity are analyzed together with several additional inputs used for multivariate training.
\begin{itemize}
    \item \textbf{Energy Scales ($H_T$, $\mathbb{E}_T^{miss}$):} We utilize the scalar sum of all jet transverse momenta, $H_T = \sum |p_{T_jet}|$, with a minimum requirement of $H_T > 450$ GeV. Additionally, we require $\mathbb{E}_T^{miss} > 10$ GeV to ensure the presence of the recoil neutrino.
    \item \textbf{Multiplicities ($N_{jets}, N_{b-jets}$):} We mandate $N_{jets} \ge 4$ and $N_{b-jets} \ge 2$. The heavy-flavor requirement is the primary veto against light-flavor $W/Z$ production and SM multijet noise. The associated production of the Vector-Like Top with a bottom quark results in a final state with multiple $b$-jets. The kinematics of the two leading $b$-jets ($p_T^{b_1}, \eta_{b_1}, p_T^{b_2}, \eta_{b_2}$) are critical for separation. Furthermore, we exploit:
%%%%%%%%%%%%%%%%%%%%%%%%%%%%%%%%%%%%%%%%%%%%%%%%%%%%%%%%%%%%%%%%%%%
\item\textbf{Angular Separation ($\Delta R_{bb}$)}: as the VLT mass increases, its boost leads to more collimated decay products, shifting the $\Delta R_{bb}$ distribution toward lower values compared to non-resonant backgrounds.
\end{itemize}

\begin{itemize}
\item \textbf{Reconstructed Mass ($M_{bjj}$):} The central resonant observable, exhibiting a Jacobian peak at the $T$ mass. The reconstruction of the VLT candidate is achieved through a global $\chi^{2}$ minimization fit. This procedure yields several high-impact variables for our multivariate analysis. The algorithm iterates through jet permutations to find the most likely $W \to jj$ and $T \to Wb$ candidates by minimizing:
\begin{equation}
    \chi^2 = \frac{(M_{jj} - M_W)^2}{\sigma_W^2} + \frac{(M_{jjb} - M_{top})^2}{\sigma_{top}^2}
\end{equation}
   \item \textbf{Reconstructed Top $p_T$ ($p_T^{T}$):} Captures the extreme boost of the VLT candidate, which is significantly higher than that of SM single-top or $t\bar{t}$ production.
\end{itemize}
\begin{table}[H]
\label{tab:III}
  \caption{Distribution of the Cumulative Signal Efficiencies for Hadronic Channel across all $m_{T}$ (2000-5000 GeV) points at $\sqrt{s}$ = 9.16 TeV.}
  \centering
  \small
  \resizebox{\textwidth}{!}{%
  \begin{tabular}{|l|r|r|r|r|r|r|r|}
  \hline
  \textbf{Selection Cuts} & \textbf{2000 GeV} & \textbf{2500 GeV} & \textbf{3000 GeV} & \textbf{3500 GeV} & \textbf{4000 GeV} & \textbf{4500 GeV} & \textbf{5000 GeV} \\
  \hline
  $|\eta_j| < 2.5$ & 0.49976 & 0.50416 & 0.49884 & 0.50292 & 0.49784 & 0.49768 & 0.50296 \\
  \hline
  $p_T^{j_1} > 120\,\mathrm{GeV}$ & 0.49656 & 0.50168 & 0.49704 & 0.50060 & 0.49608 & 0.49616 & 0.50100 \\
  \hline
  $p_T^{j_2} > 50\,\mathrm{GeV}$ & 0.49276 & 0.49876 & 0.49448 & 0.49896 & 0.49428 & 0.49436 & 0.49944 \\
  \hline
  $N_{\mathrm{jets}} \ge 4$ & 0.25232 & 0.25824 & 0.24812 & 0.25188 & 0.24996 & 0.23784 & 0.24324 \\
  \hline
  $N_{b\mathrm{jets}} \ge 2$ & 0.08516 & 0.07564 & 0.06688 & 0.05968 & 0.05884 & 0.05088 & 0.04876 \\
  \hline
  $H_T > 450\,\mathrm{GeV}$ & 0.08476 & 0.07532 & 0.06672 & 0.05964 & 0.05884 & 0.05080 & 0.04876 \\
  \hline
  $E_T^{\mathrm{miss}} > 10\,\mathrm{GeV}$ & 0.07848 & 0.07084 & 0.06284 & 0.05680 & 0.05512 & 0.04836 & 0.04632 \\
  \hline
  \textbf{$\epsilon$\%} & \textbf{7.848} & \textbf{7.084} & \textbf{6.284} & \textbf{5.680} & \textbf{5.512} & \textbf{4.836} & \textbf{4.632} \\
  \hline
  \end{tabular}%
  }
  \end{table}
%%%%%%%%%%%%%%%%%%%%%%%%%%%%%%%%%%%%%%%%%%%%%%%%%%%%%%%%%%%%%%%%%%%
As detailed in the signal-efficiency tables, the reconstruction efficiency is strongly energy-dependent. At $\sqrt{s}=9.16$ TeV, the cumulative efficiency for a 5000 GeV VLT reaches 4.632\% after the full reconstruction sequence, which is roughly three times the corresponding value at $\sqrt{s}=5.29$ TeV. This illustrates the stronger projected sensitivity at higher center-of-mass energies, where the signal is less compressed against the SM background tail.
%%%%%%%%%%%%%%%%%%%%%%%%%%%%%%%%%%%%%%%%%%%%%%%%%
\subsection{Leptonic Section} 
%%%%%%%%%%%%%%%%%%%%%%%%%%%%%%%%%%%%%%%%%%%%%%%%%
The search for a Vector-Like Singlet Top ($T$) Quark at a muon-proton collider in the leptonic final state considers the signal $\mu^- p \to \nu_\mu T \bar{b} \to \nu_\mu (Wb)\bar{b} \to \nu_\mu (\ell \nu b)\bar{b}$, which contains a charged lepton, multiple neutrinos, and heavy-flavor jets. The dominant Standard Model backgrounds are single-top production, $\mu^- p \to \nu_\mu t \bar{b} \to \nu_\mu (W^+ b) \bar{b} \to \nu_\mu (\ell^+ \nu_\ell b) \bar{b}$, and $W$+jets production, $\mu^- p \to \nu_\mu W^+ j \to \nu_\mu (\ell^+ \nu_\ell) j$. The data samples were simulated using MadGraph5 \cite{madgraph} followed by parton showering with PYTHIA 8.3 \cite{Bierlich:2022pythia}. Detector effects were then applied using DELPHES 3.5 \cite{delphes}. To isolate the multi-TeV signal, a sequential selection strategy is implemented and described in detail. The impact of each cut on the signal efficiencies across the three center-of-mass energies ($\sqrt{s}=5.29$, $6.48$, and $9.16$ TeV) is detailed in Table~\ref{IVS}.

%%%%%%%%%%%%%%%%%%%%%%%%%%%Ferynmann Diagram for Leptoni Channel%%%%%%%%%%%%%%%%%%%%%
\begin{figure}[H]
\centering
\begin{tikzpicture}
\begin{feynman}

\vertex (mu) {\(\mu^-\)};
\vertex[below=1.8cm of mu] (p) {\(p\)};

\vertex[right=2.0cm of mu] (v1);
\vertex[right=2.0cm of p] (v2);
\vertex[right=2.2cm of v1] (nu) {\(\nu_\mu\)};
\vertex[right=2.2cm of v2] (T) {\(T\)};

\vertex[right=1.4cm of T] (v3);
\vertex[above right=0.9cm and 0.9cm of v3] (e) {\(e^+\)};
\vertex[below right=0.9cm and 0.9cm of v3] (nue) {\(\nu_e\)};
\vertex[below=1.0cm of T] (b) {\(b\)};

\diagram*{
    (mu) -- [fermion] (v1) -- [fermion] (nu),
    (p) -- [fermion, edge label=\(b\)] (v2),
    (v1) -- [boson, edge label=\(W^-\)] (v2),
    (v2) -- [fermion, edge label=\(T\)] (T),
    (T) -- [fermion] (b),
    (T) -- [boson, edge label'=\(W^+\)] (v3),
    (v3) -- [fermion] (e),
    (v3) -- [fermion] (nue)
};
\end{feynman}
\end{tikzpicture}

\caption{Representative leading-order (LO) Feynman diagram for the single production of a vector-like singlet top quark in the leptonic channel at a $\mu p$ collider. The diagram highlights the underlying sub-process $\mu^- b \to \nu_\mu T$ (where the $b$ quark arises from $g \to b\bar{b}$ splitting inside the proton), followed by $T \to W^+ b$ and $W^+ \to e^+ \nu_e$.}

\label{fig:feyn_leptonic}

\end{figure}
%%%%%%%%%%%%%%%%%%%%%%%%%%%%%%%%%%%%%%%%%%%%%%%%%%%%%%%%%%%%%%

\begin{itemize}
    \item \textbf{Lepton Acceptance and $p_T^{\ell} > 150$ GeV:} The initial requirements for an isolated lepton ($|\eta| < 2.4$) maintain an efficiency of $\sim 35-37\%$. The $p_T$ threshold of $150$ GeV is specifically designed to suppress the soft leptonic tail from Background 2. Interestingly, the efficiency of the $p_T$ cut increases with $M_T$; for instance, at $5.29$ TeV, efficiency rises from $32.2\%$ for $M_T=2000$ GeV to $34.8\%$ for $M_T=5000$ GeV. Physically, as the VLT mass increases, its decay products are naturally more energetic, allowing a higher fraction to surpass the hard trigger threshold.
    
    \item \textbf{Jet Multiplicity and B-Tagging:} We mandate $N_{jets} \ge 2$ ($p_T > 30$ GeV) and exactly one $b$-tagged jet ($p_T > 80$ GeV). This stage represents the most significant efficiency "tax." At $\sqrt{s}=9.16$ TeV, the efficiency drops from $\sim 28\%$ to $\sim 10\%$ for a $4000$ GeV mass. This drop is due to the extreme Lorentz boost of the $T$ quark; as its products become collimated, the $b$-jet often merges with the lepton or neutrinos, or fails the isolation criteria for high-mass resonances \cite{Rehermann:2010top}.

    \item \textbf{Missing Energy ($MET > 80$ GeV):} Since the signal production vertex and $W$ decay both involve neutrinos, $MET$ is a crucial discriminator. The efficiency remains stable across mass points ($\sim 9-10\%$ cumulative), effectively filtering out QCD events with instrumental $MET$.
   
    \item \textbf{Angular Collimation ($\Delta R(\ell, b) > 1.5$):} This cut is the primary veto for the SM single-top background. As VLT mass increases, the efficiency for this cut consistently decreases—dropping from $9.97\%$ to $7.26\%$ at $6.48$ TeV. This trend is a signature of the "boosted" regime where decay products are forced into a narrow cone, a physical effect more pronounced at higher masses.
    \end{itemize}
%%%%%%%%%%%%%%%%%%%%%%%%%%%%%%%%%%%%%%%%%%%%%%%%%%%%%%%%%%%%%%%%%%%
  \begin{table}[H]
  \label{tab:IV}
  \caption{Cumulative signal efficiencies for the leptonic channel across all benchmark masses at $\sqrt{s}=5.29$ TeV.}
  \centering
  \small
  \resizebox{\textwidth}{!}{%
  \begin{tabular}{|l|r|r|r|r|r|r|r|}
  \hline
  \textbf{Selection Cuts} & \textbf{2000 (GeV)} & \textbf{2500 (GeV)} & \textbf{3000 (GeV)} & \textbf{3500 (GeV)} & \textbf{4000 (GeV)} & \textbf{4500 (GeV)} & \textbf{5000 (GeV)} \\
  \hline
  $N_{\ell}$ $\ge$ 1 ($P_{T}$ $>$ 10) & 0.37504 & 0.37572 & 0.36992 & 0.36660 & 0.36452 & 0.36688 & 0.36156 \\
  \hline
 $\eta_{\ell}$  $<$ 2.4 & 0.37460 & 0.37524 & 0.36964 & 0.36628 & 0.36408 & 0.36648 & 0.36092 \\
  \hline
   $p_T^{\ell} > 150$ GeV & 0.32204 & 0.33828 & 0.34156 & 0.34408 & 0.34656 & 0.35172 & 0.34860 \\
  \hline
 $N_{jets} \ge 2$ ($p_T > 30$ GeV) & 0.23708 & 0.24880 & 0.25024 & 0.24572 & 0.24600 & 0.23456 & 0.17488 \\
  \hline
   $ b_{jet} \equiv 1$ ($P_{T}$ $>$ 80) & 0.10820 & 0.10576 & 0.09420 & 0.08796 & 0.08452 & 0.07280 & 0.04544 \\
  \hline
  MET $>$ 80 GeV & 0.09628 & 0.09752 & 0.08816 & 0.08248 & 0.07960 & 0.06840 & 0.04308 \\
  \hline
 $\Delta R(\ell, b) > 1.5$ & 0.09180 & 0.09320 & 0.08332 & 0.07828 & 0.07532 & 0.06432 & 0.04248 \\
  \hline
  \textbf{$\epsilon$\%} & \textbf{9.180} & \textbf{9.320} & \textbf{8.332} & \textbf{7.828} & \textbf{7.532} & \textbf{6.432} & \textbf{4.248} \\
  \hline
  \end{tabular}%
  }
  \end{table}
%%%%%%%%%%%%%%%%%%%%%%%%%%%%%%%%%%%%%%%%%%%%%%%%%%%%%%%%%%%%%%%%%%%
 \begin{table}[H]
 \label{tab:V}
  \caption{Cumulative signal efficiencies for the leptonic channel across all benchmark masses at $\sqrt{s}=6.48$ TeV.}
  \centering
  \small
  \resizebox{\textwidth}{!}{%
  \begin{tabular}{|l|r|r|r|r|r|r|r|}
  \hline
  \textbf{Selection Cuts} & \textbf{2000 (GeV)} & \textbf{2500 (GeV)} & \textbf{3000 (GeV)} & \textbf{3500 (GeV)} & \textbf{4000 (GeV)} & \textbf{4500 (GeV)} & \textbf{5000 (GeV)} \\
  \hline
  $N_{\ell}$) $\ge$ 1 ($P_{T}$ $>$ 10) & 0.37292 & 0.36832 & 0.36876 & 0.36696 & 0.35872 & 0.36312 & 0.36208 \\
  \hline
 $\eta_{\ell}$ $<$ 2.4 & 0.37204 & 0.36768 & 0.36820 & 0.36664 & 0.35864 & 0.36260 & 0.36156 \\
  \hline
  $p_T^{\ell} > 150$ GeV & 0.32404 & 0.33244 & 0.34112 & 0.34648 & 0.34392 & 0.34876 & 0.34900 \\
  \hline
$N_{jets} \ge 2$ ($p_T > 30$ GeV) & 0.25128 & 0.26008 & 0.26572 & 0.26648 & 0.26324 & 0.26296 & 0.25860 \\
  \hline
   $ b_{jet} \equiv 1$ ($P_{T}$ $>$ 80) & 0.11496 & 0.10888 & 0.10296 & 0.09768 & 0.09128 & 0.08680 & 0.08092 \\
  \hline
  MET $>$ 80 GeV & 0.10408 & 0.10088 & 0.09552 & 0.09184 & 0.08664 & 0.08228 & 0.07708 \\
  \hline
  $\Delta R(\ell, b) > 1.5$ & 0.09972 & 0.09696 & 0.09064 & 0.08716 & 0.08096 & 0.07692 & 0.07260 \\
  \hline
  \textbf{$\epsilon$\%} & \textbf{9.972} & \textbf{9.696} & \textbf{9.064} & \textbf{8.716} & \textbf{8.096} & \textbf{7.692} & \textbf{7.260} \\
  \hline
  \end{tabular}%
  }
  \end{table}
%%%%%%%%%%%%%%%%%%%%%%%%%%%%%%%%%%%%%%%%%%%%%%%%%%%%%%%%%%%%%%%%%%%
  \begin{table}[H]
  \label{tab:VI}
  \caption{Cumulative signal efficiencies for the leptonic channel across all benchmark masses at $\sqrt{s}=9.16$ TeV.}
  \centering
  \small
  \resizebox{\textwidth}{!}{%
  \begin{tabular}{|l|r|r|r|r|r|r|r|}
  \hline
  \textbf{Selection Cuts} & \textbf{2000 (GeV)} & \textbf{2500 (GeV)} & \textbf{3000 (GeV)} & \textbf{3500 (GeV)} & \textbf{4000 (GeV)} & \textbf{4500 (GeV)} & \textbf{5000 (GeV)} \\
  \hline
 $N_{\ell}$ $\ge$ 1 ($P_{T}$ $>$ 10) & 0.35056 & 0.35216 & 0.35600 & 0.35132 & 0.35616 & 0.35552 & 0.35368 \\
  \hline
  $\eta_{\ell}$ $<$ 2.4 & 0.34880 & 0.35052 & 0.35488 & 0.35020 & 0.35528 & 0.35468 & 0.35292 \\
  \hline
   $p_T^{\ell} > 150$ GeV & 0.30532 & 0.32152 & 0.33252 & 0.33240 & 0.34040 & 0.34264 & 0.34228 \\
  \hline
 $N_{jets} \ge 2$ ($p_T > 30$ GeV) & 0.24812 & 0.26388 & 0.27676 & 0.27504 & 0.27980 & 0.28236 & 0.27692 \\
  \hline
   $ b_{jet} \equiv 1$ ($P_{T}$ $>$ 80) & 0.11176 & 0.11188 & 0.10968 & 0.10152 & 0.09968 & 0.09936 & 0.09284 \\
  \hline
  MET $>$ 80 GeV & 0.10196 & 0.10328 & 0.10316 & 0.09660 & 0.09472 & 0.09452 & 0.08940 \\
  \hline
  $\Delta R(\ell, b) > 1.5$ & 0.09732 & 0.09724 & 0.09736 & 0.09024 & 0.08796 & 0.08820 & 0.08268 \\
  \hline
  \textbf{$\epsilon$\%} & \textbf{9.732} & \textbf{9.724} & \textbf{9.736} & \textbf{9.024} & \textbf{8.796} & \textbf{8.820} & \textbf{8.268} \\
  \hline
  \end{tabular}%
  }
  \end{table}
%%%%%%%%%%%%%%%%%%%%%%%%%%%%%%%%%%%%%%%%%%%%%%%%%%%%%%%%%%%%%%%%%%%
The multivariate analysis uses eight physically motivated observables to separate the signal from the SM backgrounds. In particular:
\begin{itemize}
    \item \textbf{Reconstructed Mass ($M_{b\ell\nu}$):} This is the primary resonant variable. By solving for the neutrino $p_z$ using the $W$-mass constraint, we reconstruct the parent top-partner mass. Signal events peak at $M_T$, while non-resonant $W$+jets events form a broad continuum, allowing the MVA to achieve strong separation.
    
    \item \textbf{Transverse Mass ($M_T(\ell, MET)$):} For the $W$+jets background, this variable is bounded by the $W$-mass Jacobian peak ($\sim 80$ GeV). Since the $W$ in our signal originates from a multi-TeV $T$ decay, its transverse mass extends significantly higher, providing a clean separation boundary.
    
    \item \textbf{Helicity Angle ($\cos\theta^*$):} This angular observable probes the polarization of the $W$ boson. Because the VLT singlet model involves a different chiral coupling than the SM top, the $\cos\theta^*$ distribution provides an additional discriminant once the signal and background shapes are exploited simultaneously by the multivariate classifier. It can be calculated using:
    \begin{equation}
    \cos\theta^* = \frac{\vec{p}_{\ell}^{*} \cdot \vec{p}_{W}}{|\vec{p}_{\ell}^{*}| |\vec{p}_{W}|}
\end{equation}
    
    \item \textbf{$p_T$ Ratio and MET:} The ratio of lepton $p_T$ to $MET$ or jet $p_T$ measures the kinematic balance of the system. In VLT decays, the products follow a characteristic energy-sharing profile, whereas in $W+j$ events the jet $p_T$ is often asymmetric relative to the lepton, making the ratio a useful discriminant.
\end{itemize}

\begin{figure}[H]
    \centering
    \includegraphics[height=9.5cm,width=10.5cm]{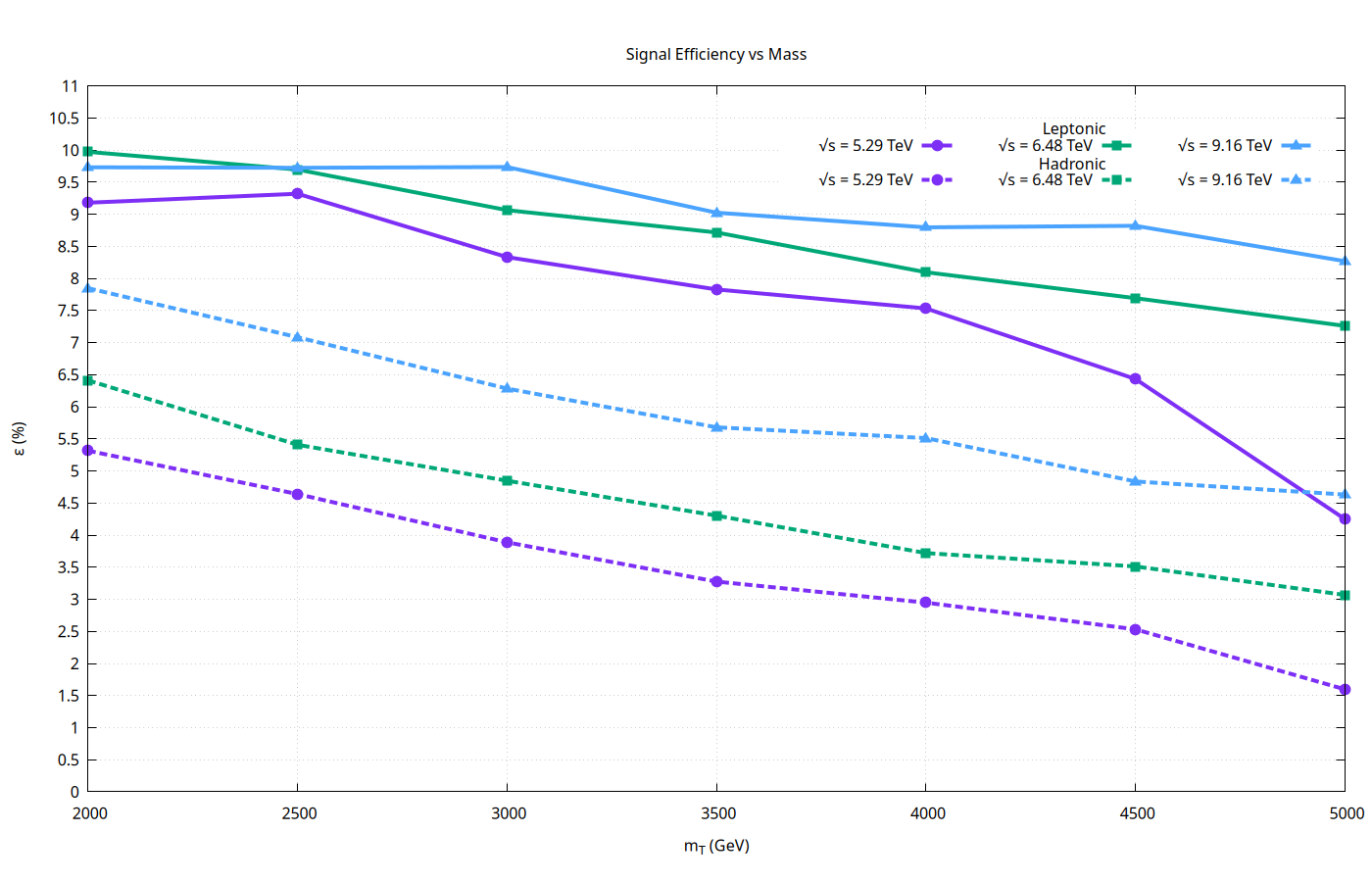}
    \caption{Hadronic and leptonic signal efficiencies $\epsilon$ (\%) as functions of $m_T$ at $\sqrt{s}=5.29$, $6.48$, and $9.16$~TeV.  The efficiency decreases toward higher masses in both channels.  The leptonic channel remains comparatively stable across the scan, whereas the hadronic efficiency falls more visibly at the highest masses because the resolved selection loses acceptance as the decay products become increasingly collimated.}
    \label{fig:eff-had}
  \end{figure}

%%%%%%%%%%%%%%%%%%%%%%%%%%%%%%%%%%%%%%%%%%%%%%%%%%%%%%%%%%%%%%%%%%%
The signal efficiencies are taken together and displayed in Fig.~\ref{fig:eff-had}. We observe that the leptonic mode provides a cleaner signature and a useful cross-check, which enhances the overall search sensitivity of the future muon-proton collider \cite{Ahmed:2025up, Sirunyan:2019sga}.
%%%%%%%%%%%%%%%%%%%%%%%%%%%%%%%%%%%%%%%%%%%%%%%%%%%%%%%%%%%%
  % Hadronic efficiency and Leptonic efficiency
  
\section{Advanced Machine Learning Techniques for the Feasibility Study}
\label{sec:mva}
\subsection{Signal Extraction Through Multivariate Analysis}
%%%%%%%%%%%%%%%%%%%%%%%%%%%%%%%%%%%%%%%%%%%%%%%%%%%%%%%%%%%%%%%%%%%%%%%%

To classify and separate signal events from the background ensemble, we use four TMVA classifiers (machine-learning methods that are now widely used in LHC physics \cite{Guest:2018bsm}): Boosted Decision Trees (BDT), Gradient Boosted Decision Trees (BDTG), Multilayer Perceptrons (MLP), and Likelihood \cite{Hocker:2007ht}.  Multivariate techniques offer a significant advantage over traditional cut-based methods because they can exploit high-dimensional and non-linear correlations among kinematic observables that are only partially accessible in a sequential rectangular-cut analysis.

The multivariate analysis is used here as a physics-guided signal-extraction step, not as a replacement for the event interpretation.  The input observables are chosen because they correspond to identifiable features of the $T\to Wb$ topology: a resonant mass structure, hard recoil against the spectator system, $b$-jet activity, missing transverse momentum, and angular separation patterns.  This makes the classifier output interpretable: a signal-like event is not accepted because of an opaque numerical score alone, but because several physically motivated handles point consistently toward a heavy resonant topology.

TMVA classifiers such as BDTs employ a tree-based architecture typically consisting of several hundred individual decision trees. In this study, we utilize the ``AdaBoost'' and ``GradBoost'' algorithms to iteratively assign higher weights to misclassified events, thereby improving the separation power in regions where signal and background distributions overlap significantly.

Similarly, the Multilayer Perceptron (MLP) is an Artificial Neural Network (ANN) characterized by a multi-layered architecture consisting of input, hidden, and output layers trained by back-propagation \cite{Rumelhart:1986}. Each neuron in the hidden layer performs a non-linear transformation of the physical variables using an ``Activation Function'' (such as Sigmoid or ReLU). This process allows the network to check for complex correlations and model the underlying patterns in the ensemble distribution. After successful training, the MLP provides a distribution score between 0 and 1, facilitating a visual and statistical classification between signal and background.

The Likelihood classifier is a simpler approach compared to BDTs and MLP, as it operates on the ``Naive Bayes'' assumption that input variables are independent. This method constructs probability density functions (PDFs) by generating histograms for signal ($f_s$) and background ($f_B$) for each variable. In this setup, the signal and background likelihoods are built as:
\begin{equation}
L_s(x) = \prod_{i=1}^n f_{s,i}(x_i), \quad L_B(x) = \prod_{i=1}^n f_{B,i}(x_i)
\end{equation}

Finally, the projected likelihood computes the likelihood ratio, which serves as the final discriminant:
\begin{equation}
y_{LH}(x) = \frac{L_s(x)}{L_s(x) + L_B(x)}
\end{equation}

The resulting distribution scores for both the Likelihood and the tree-based classifiers are presented between $-1$ and $1$.  The projected-likelihood classifier performs less well here because the training variables are significantly correlated, so it is retained mainly as a reference method rather than as the optimal working point.  This behavior does not indicate over-training.

All classifiers are trained using a set of input variables for both the hadronic and leptonic channels that provide the most separation power and discrimination between signal and background events. These variables include kinematic properties of the reconstructed objects, and b-tagging information. The features that are used as inputs to classifiers are listed in Table~\ref{IVS}. 
%%%%%%%%%%%%%%%%%%%%%%%%%%%%%%%%%%%%%%%%%%%%%%%%%%%%%%%%%%%%%%%%%%%%%%%%%%%%%%%%%%%%%5%%%%%
 
 \begin{table}[htbp]
  \caption{Input variables used for the machine-learning classifiers in the hadronic and leptonic analyses at $\sqrt{s}=5.29$, $6.48$, and $9.16$ TeV.}
  \label{IVS}
  \centering
 \renewcommand{\arraystretch}{1.2}
  \resizebox{\textwidth}{!}{%
  \begin{tabular}{|c|c|c|c|c|}
  \hline
  \multicolumn{4}{|c|}{\textbf{Distribution of Input Variables}} \\
  \hline
  Leptonic & Description & Hadronic & Description \\
  \hline
  $M_{b\ell\nu}$ & Reconstructed Mass of Vector-Like Top Quark & $E_T^{miss}$ & Missing Transverse Energy \\
  \hline
  $\rho_{T l}^{\pm}$ & Transverse Momenta of Leptons & $H_T$ & Sum of Transverse Momenta of all the jets ( $\sum |p_{T_jet}|$) \\
  \hline
  $\Delta R(\ell, b)$ & Angular Correlation of Leptons and b-tagged jets & $\Delta R_{b_{j1},\,b_{j2}}$ & Angular Separation between leading and sub-leading b-tagged jets \\
  \hline
  $\rho_{T bjet}$ & Transverse Momenta of b-tagged jets & $M_{bjj}$ & Reconstructed Mass of Vector-Like Top Quark\\
  \hline
   {$\cos\theta^*$} & {Helicity Angle of the decay products $(\ell, W)$}   & $\rho_{T^T}$ & Transverse Momentum of Vector-like Top Quark \\
  \hline
  $E_T^{miss}$ & Missing Transverse Energy & $\rho_{T j1}$ & Transverse Momentum of Leading jet \\
  \hline
  $M_T(\ell, MET)$ & Transverse Mass of W-Boson & $\rho_{T j2}$ & Transverse Momentum of sub-leading jet \\
  \hline
  $\frac{p_T^\ell}{p_T^b}$ & Ratio of transverse momenta of leptons and b-jets & $\rho_{T bjet1}$ & Transverse Momentum of Leading b-tagged jet \\
  \hline
  &  & $\rho_{T bjet2}$ & Transverse Momentum of sub-leading b-tagged jet \\
  \hline
  \end{tabular}%
  }
  \end{table}
%%%%%%%%%%%%%%%%%%%%%%%%%%%%%%%%%%%%%%%%%%%%%%%%%%%%%%%%%%%%%%%%%%%%%%%% 

The signal and background samples are divided into statistically independent training and test subsets.  To monitor possible over-training we use the Kolmogorov--Smirnov (KS) test, which compares the cumulative distribution functions of the training and test responses for each classifier.  In practice, it checks whether both samples remain statistically compatible up to the expected fluctuations.  The test statistic is

\begin{equation}
D_{n,m} = \sup_{x} \left| F_{1,n}(x) - F_{2,n}(x) \right|
\end{equation}

and we require a $p$ value greater than $0.05$ to regard the training and test samples as statistically consistent.  Otherwise, the classifier would be considered over-trained.  The corresponding over-training checks are shown in Figs. 4 and Fig. 5 for the hadronic and leptonic analyses, respectively.

 \medskip
 %\vspace{0.6em}
%%%%%%%%%%%%%%%%%%%%%%%%%%%%%%%%%%%%%%%%%%%%%%%%%%%%%%
  \begin{figure}[H]
      \centering
      \begin{minipage}{0.4\textwidth}
        \centering\includegraphics[width=\textwidth]{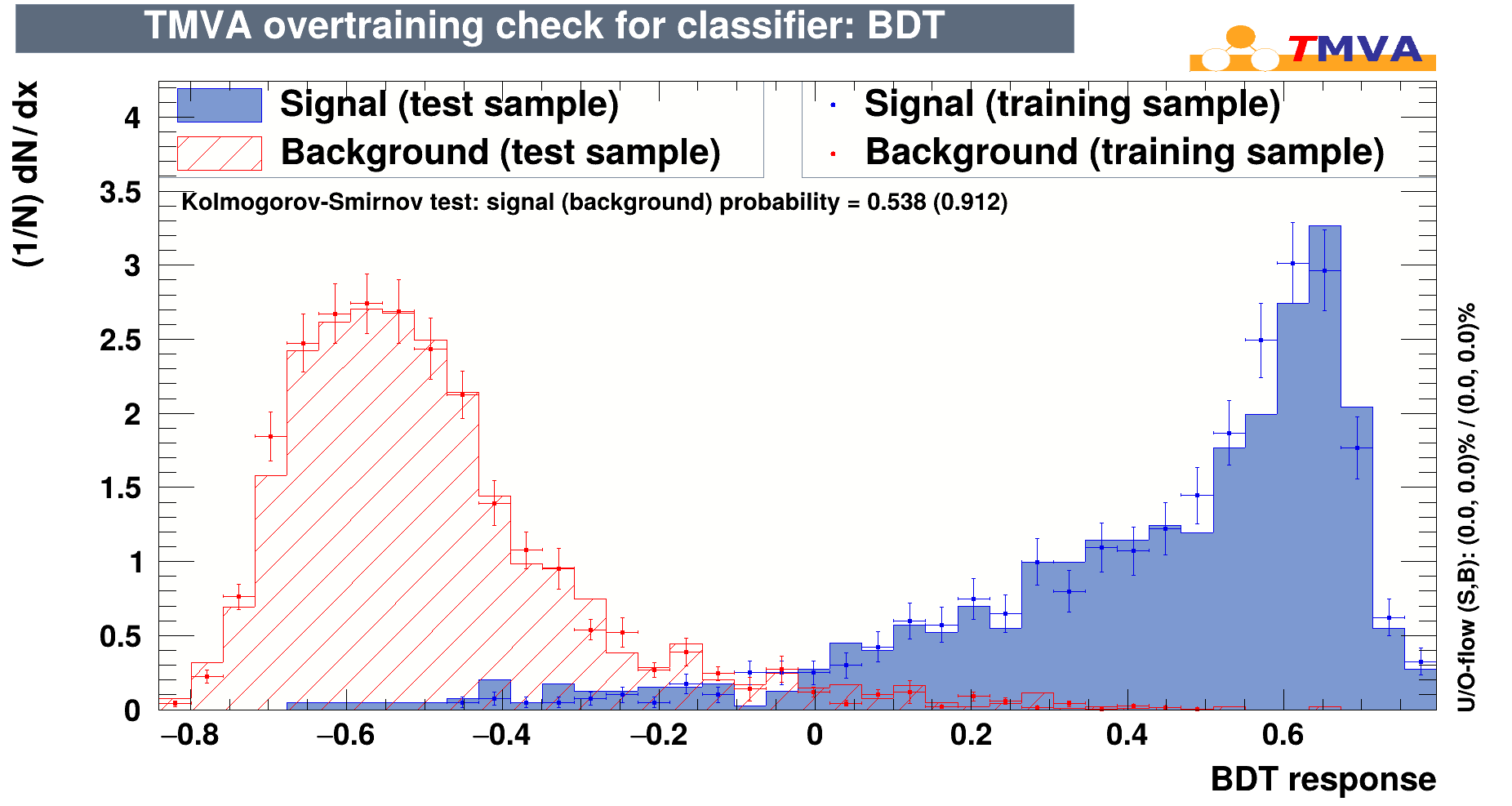}
      \end{minipage}
      \begin{minipage}{0.4\textwidth}
        \centering\includegraphics[width=\textwidth]{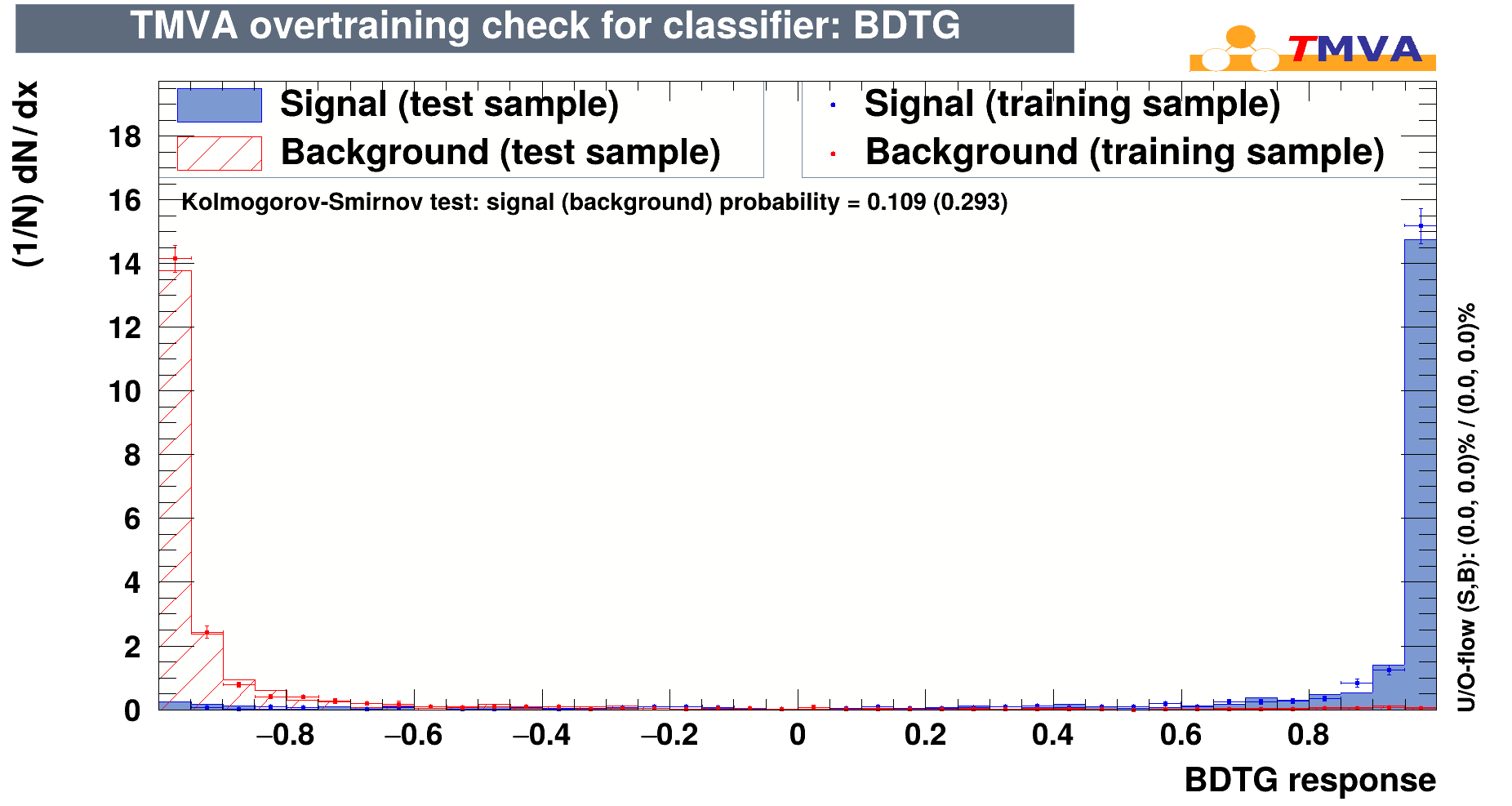}
      \end{minipage}
      \begin{minipage}{0.4\textwidth}
        \centering\includegraphics[width=\textwidth]{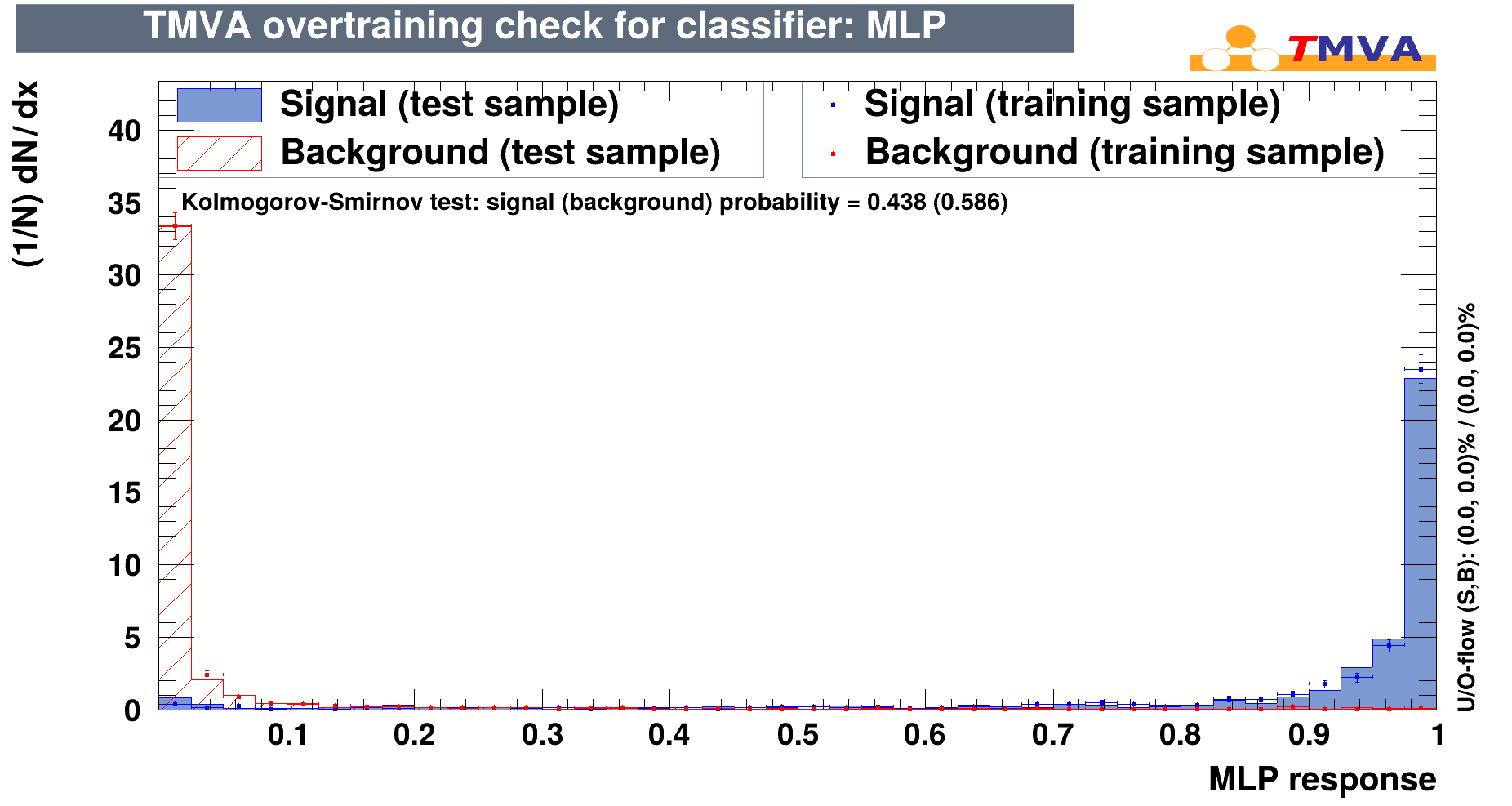}
      \end{minipage}
      \begin{minipage}{0.4\textwidth}
        \centering\includegraphics[width=\textwidth]{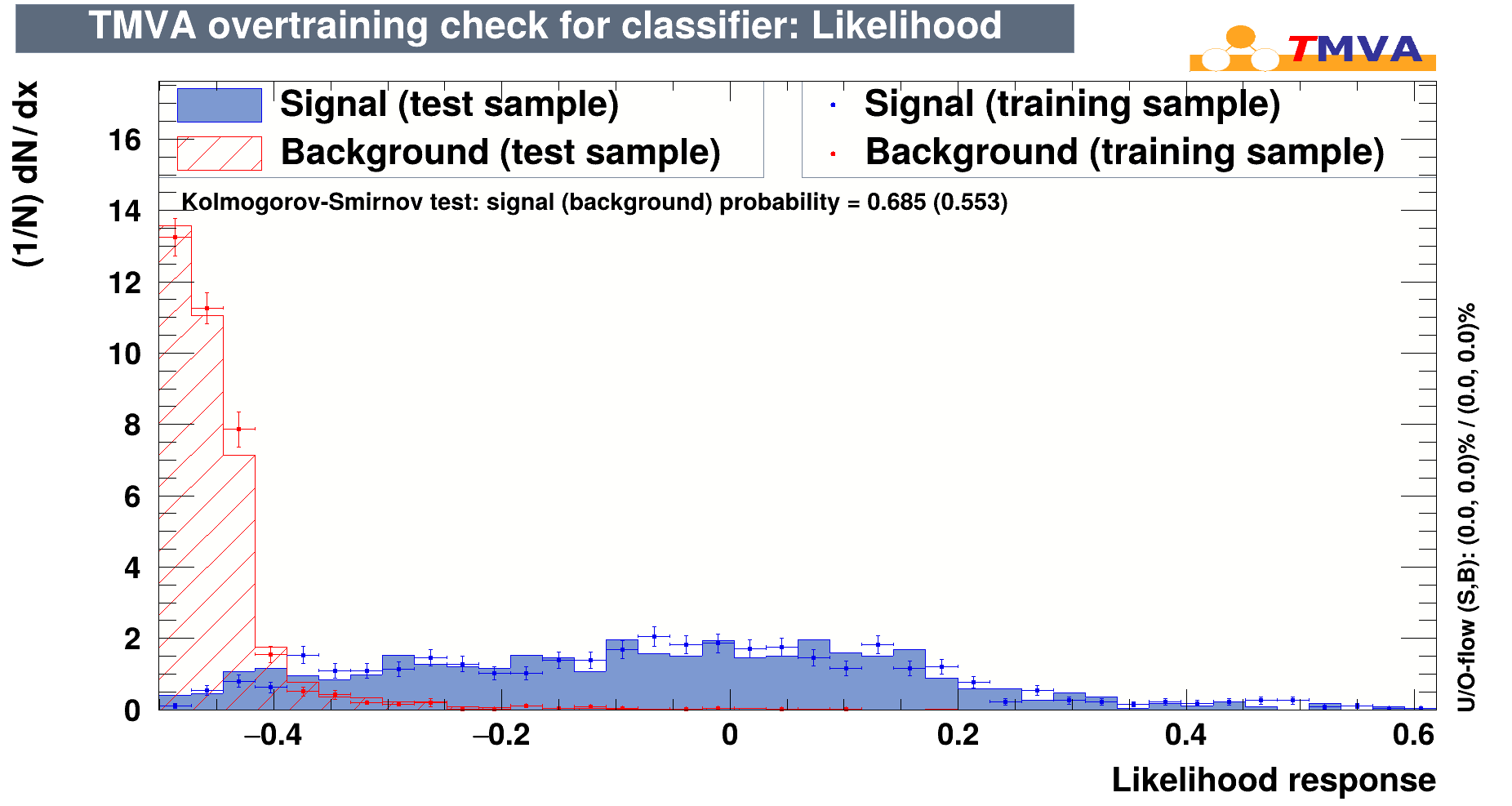}
      \end{minipage}
      \caption{Output-score distributions and over-training checks for the hadronic classifiers after multivariate training at $\sqrt{s}=9.16$~TeV for $m_T=3000$~GeV.}
      \label{fig:4}
  \end{figure}

  \begin{figure}[H]
      \centering
      \begin{minipage}{0.4\textwidth}
        \centering\includegraphics[width=\textwidth]{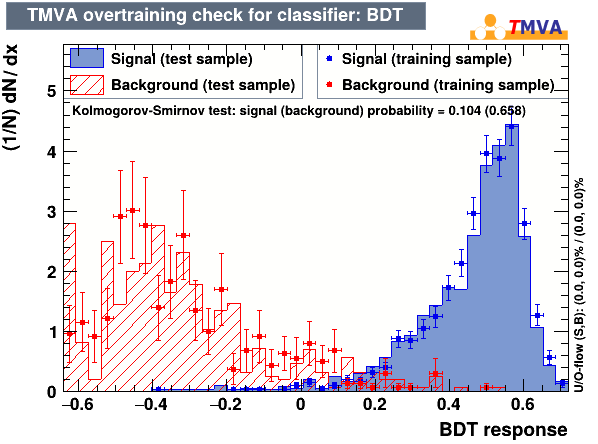}
      \end{minipage}
      \begin{minipage}{0.4\textwidth}
        \centering\includegraphics[width=\textwidth]{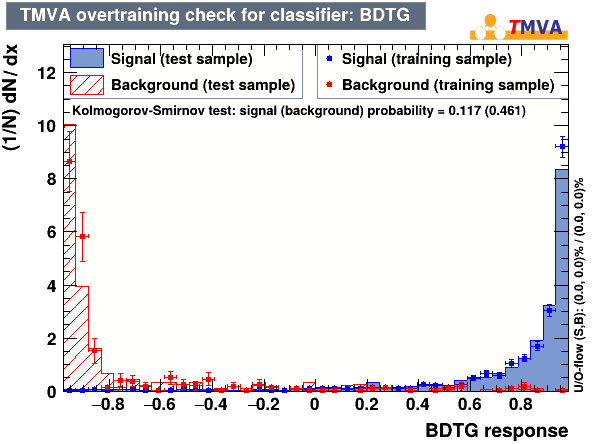}
      \end{minipage}
      \begin{minipage}{0.4\textwidth}
        \centering\includegraphics[width=\textwidth]{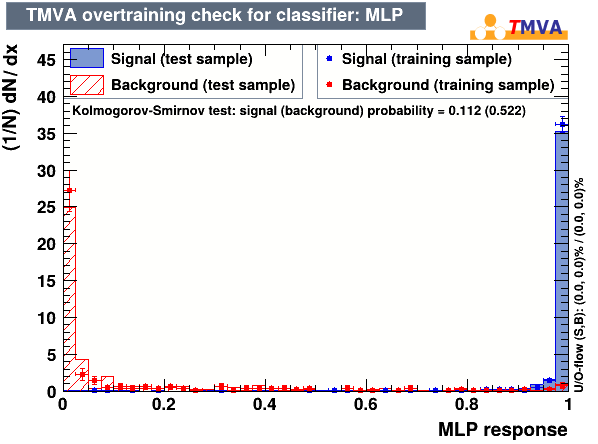}
      \end{minipage}
      \begin{minipage}{0.4\textwidth}
        \centering\includegraphics[width=\textwidth]{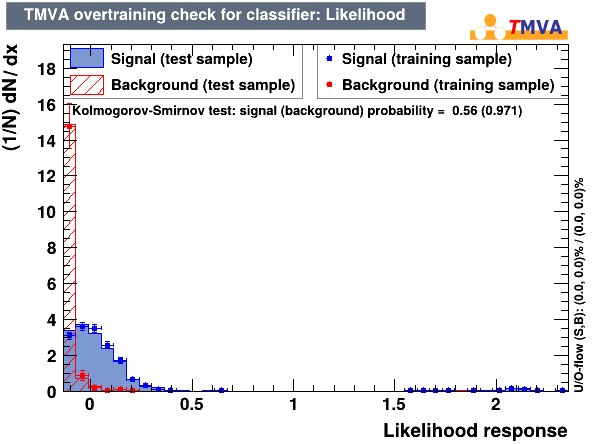}
      \end{minipage}
      \caption{Output-score distributions and over-training checks for the leptonic classifiers after multivariate training at $\sqrt{s}=9.16$~TeV for $m_T=3000$~GeV.}
      \label{fig:5}
  \end{figure}

  Figures 4 and 5 show no evidence of over-training, with $p>0.05$ for all classifiers.  The agreement between training and test samples indicates stable generalization, while the broader projected-likelihood response again reflects the stronger correlations among the input variables listed in Table~\ref{IVS}.  Hyper-parameters such as the learning rate, tree depth, number of trees, NCycles, hidden-layer structure, PDF interpolation, TestRate, and Shrinkage are tuned manually and then kept fixed across the mass scan within each analysis channel.  Their performance is evaluated through the signal-efficiency versus background-rejection tradeoff and the area under the ROC curve (AUC), summarized in Figures 6 and 7.
  
%%%%%%%%%%%%%%%%%%%%%%%%%%%%%%%%%%%%%%%%%%%%%%%%%%%%%%
 \begin{figure}[H]
    \centering
    \includegraphics[width=\linewidth]{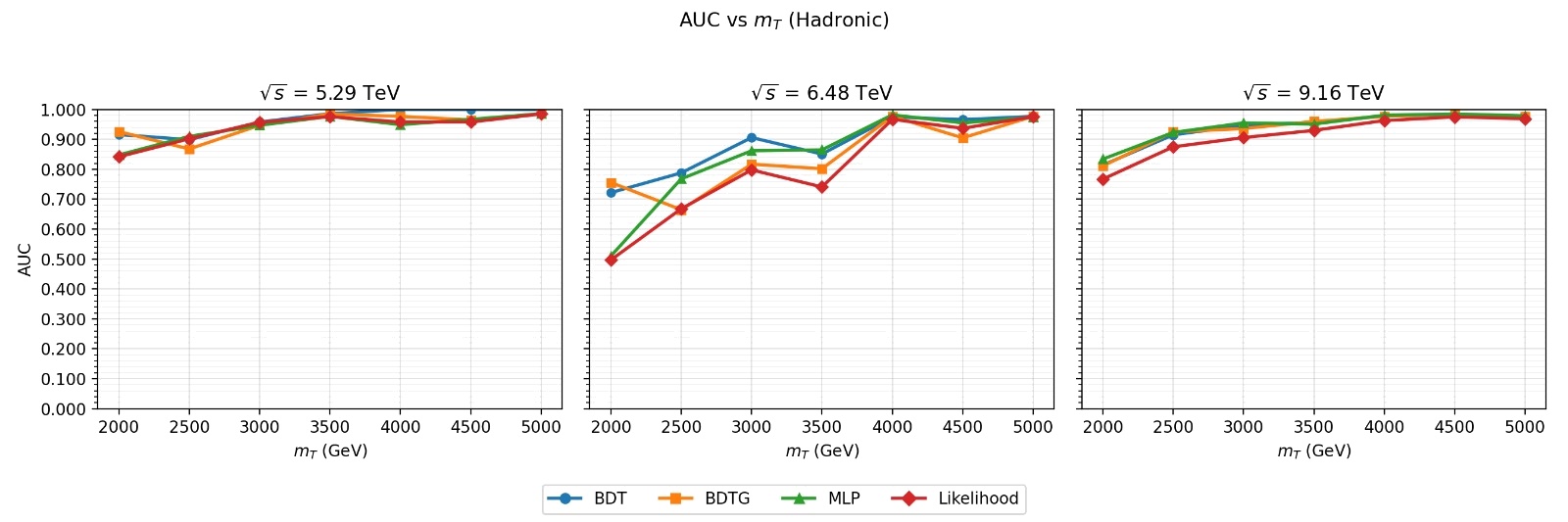}
    \caption{Comparison of the AUC as a function of $m_T$ in the hadronic analysis for the chosen classifiers (BDT, BDTG, MLP, and Likelihood) at $\sqrt{s}=5.29$, $6.48$, and $9.16$~TeV.  The curves show stable multivariate performance across the scan, with the strongest separation obtained at $\sqrt{s}=9.16$~TeV.}
    \label{fig:AUC-had}
  \end{figure}
  %%%%%%%%%%%%%%%%%%%%%%%%%%%%%%%%%%%%%%%%%%%%%%%%%%%%%%%%%%%%%%%%%%
  
  The tuned models are then applied separately to the datasets at $\sqrt{s}=5.29$, $6.48$, and $9.16$~TeV to assess how reliably the classifier performance extrapolates across the benchmark scan in both the hadronic and leptonic analyses.
  
 \begin{figure}[H]
    \centering
    \includegraphics[width=\linewidth]{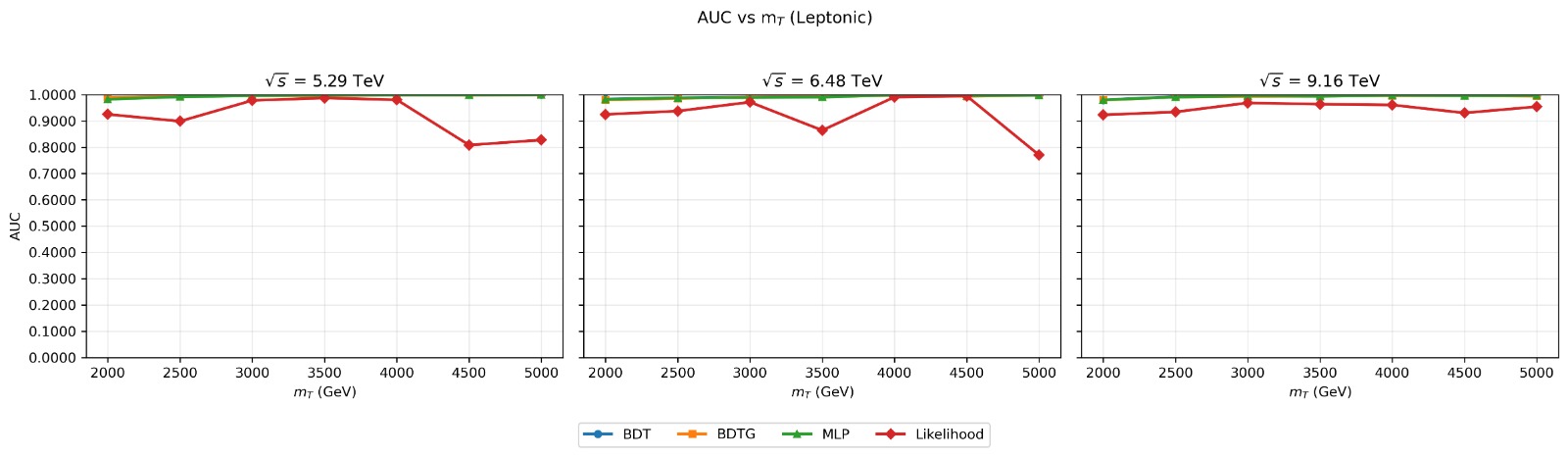}
    \caption{Comparison of the AUC as a function of $m_T$ in the leptonic analysis for the chosen classifiers (BDT, BDTG, MLP, and Likelihood) at $\sqrt{s}=5.29$, $6.48$, and $9.16$~TeV.  The leptonic classifiers also perform best at $\sqrt{s}=9.16$~TeV while remaining stable across the scanned mass range.}
    \label{fig:AUC-lep}
  \end{figure}
%%%%%%%%%%%%%%%%%%%%%%%%%%%%%%%%%%%%%%%%%%%%%%%%%%%%%%%%%%%%%%%%%%%
\subsection{The Reconstruction of Mass Points (Hadronic)}
%%%%%%%%%%%%%%%%%%%%%%%%%%%%%%%%%%%%%%%%%%%%%%%%%%%%%%%%%%%%%%%%%%%%%%%%
The primary objective of the hadronic analysis is the precise reconstruction of the vector-like $T$ quark mass from its decay products. In the hadronic channel, the $T$ quark decays via $T \to Wb$, with the $W$ boson subsequently decaying into two light-flavor jets ($W \to jj$). Consequently, the mass of the top partner is reconstructed using the invariant mass of the three-jet system, denoted as $M_{bjj}$. We utilize the anti-$k_t$ jet clustering algorithm with a radius parameter of $R = 0.4$ to define the jets, as implemented in the Delphes fast-simulation framework \cite{Cacciari:2008gp, delphes}.

\begin{figure}[H]
  \centering
  \begin{minipage}[b]{0.28\textwidth}
    \centering
    \includegraphics[width=\textwidth]{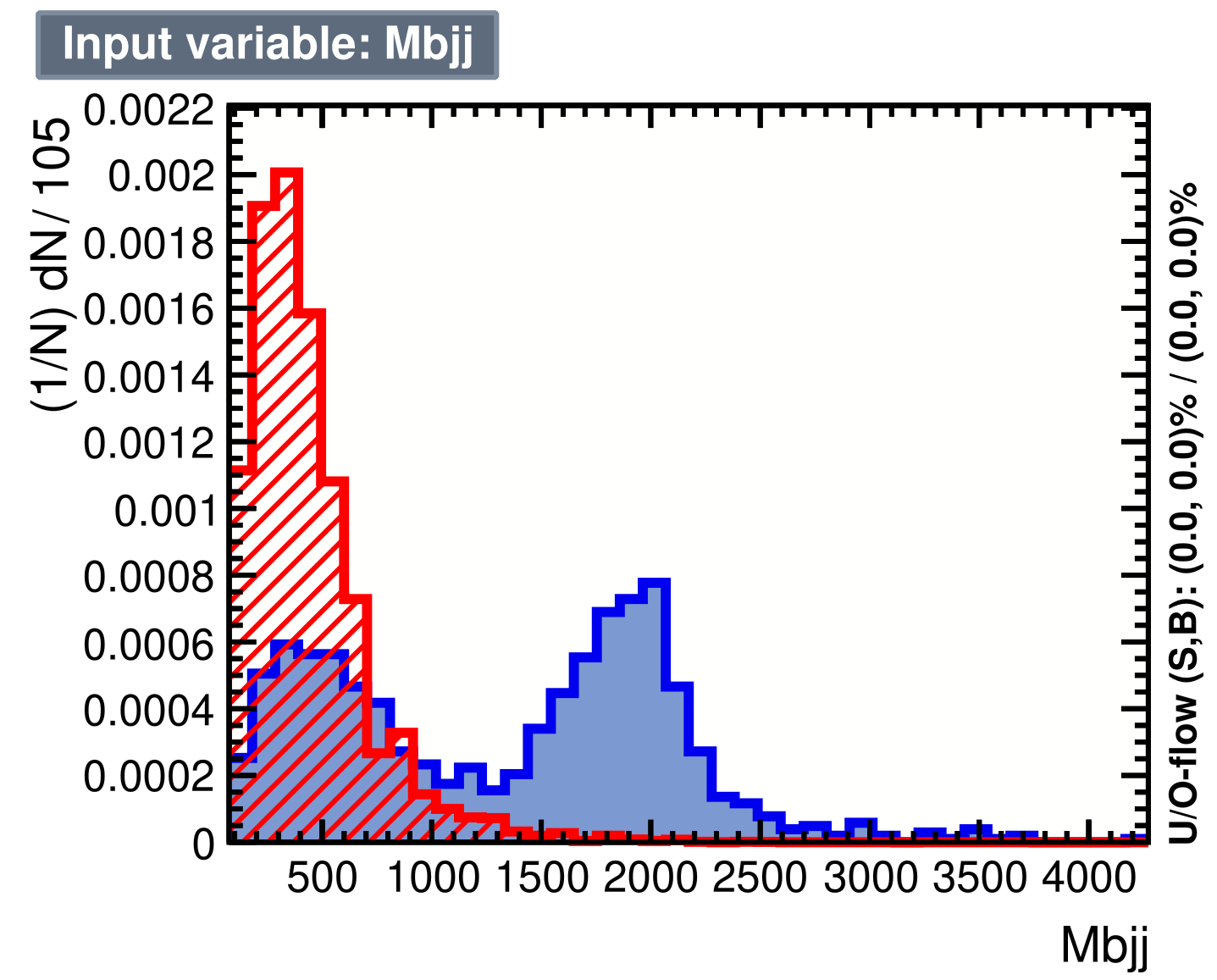}
    \par\smallskip (a)
  \end{minipage}\hfill
  \begin{minipage}[b]{0.28\textwidth}
    \centering
    \includegraphics[width=\textwidth]{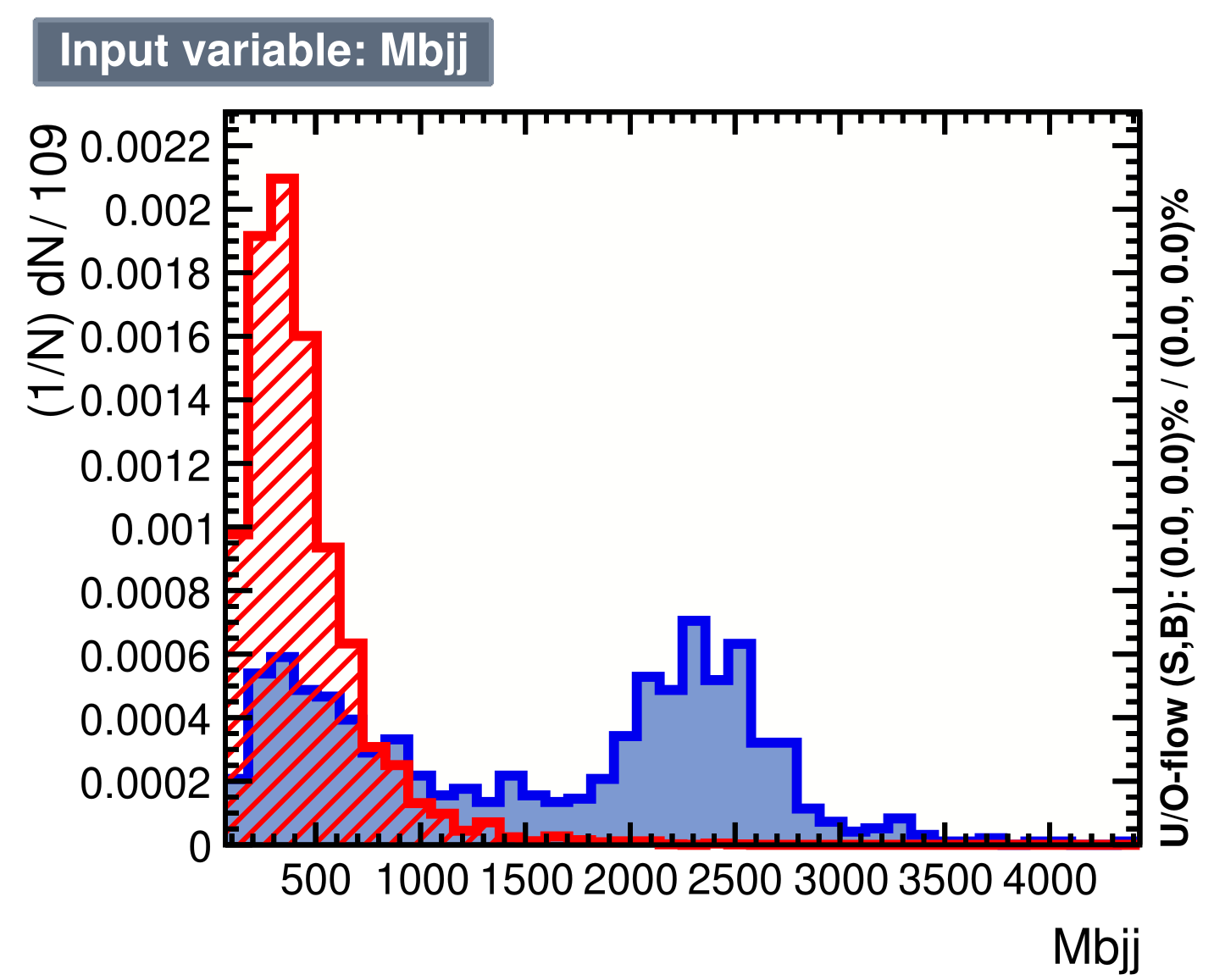}
    \par\smallskip (b)
  \end{minipage}\hfill
  \begin{minipage}[b]{0.28\textwidth}
    \centering
    \includegraphics[width=\textwidth]{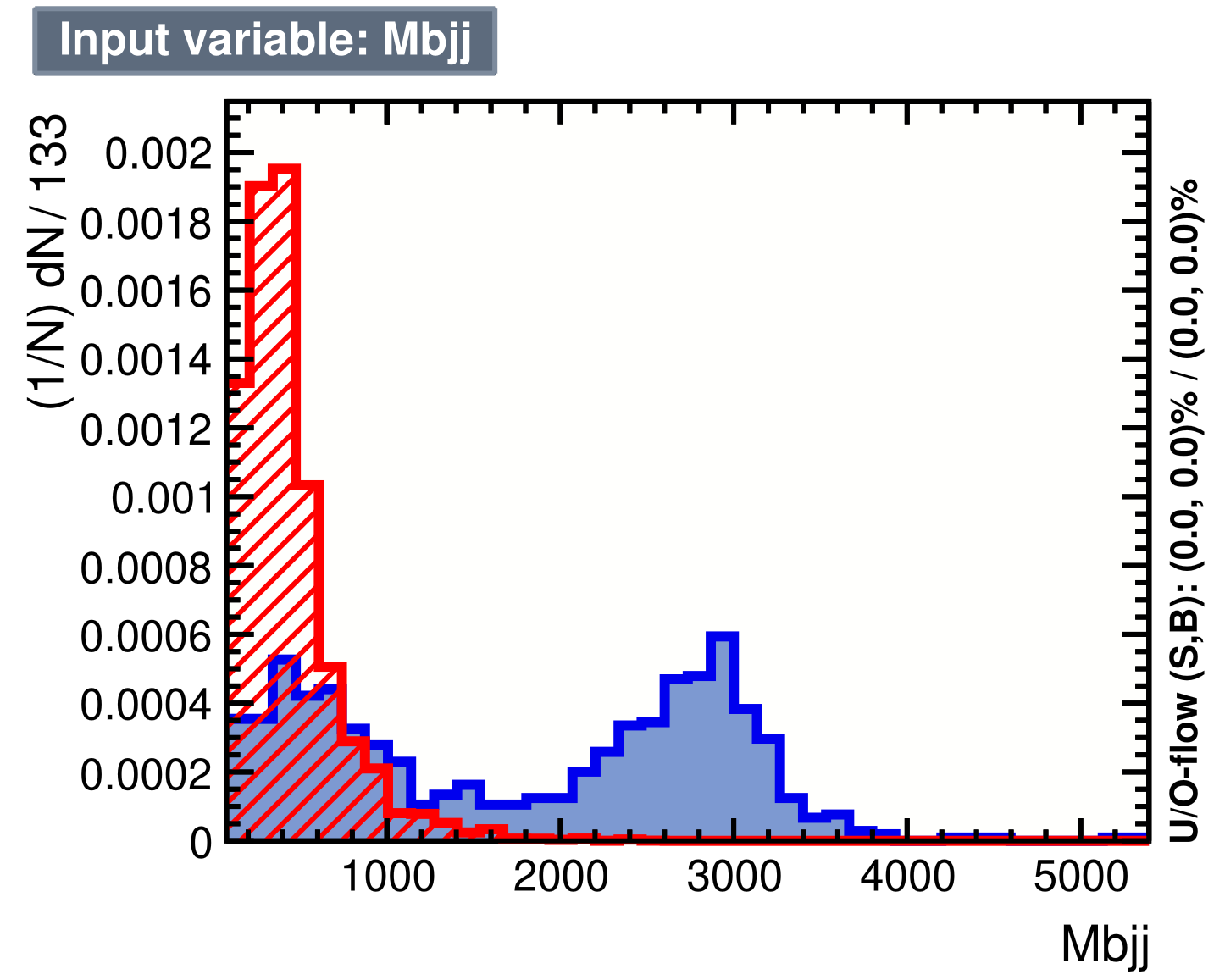}
    \par\smallskip (c)
  \end{minipage}

  \vspace{0.5em} % vertical space between rows

  \begin{minipage}[b]{0.28\textwidth}
    \centering
    \includegraphics[width=\textwidth]{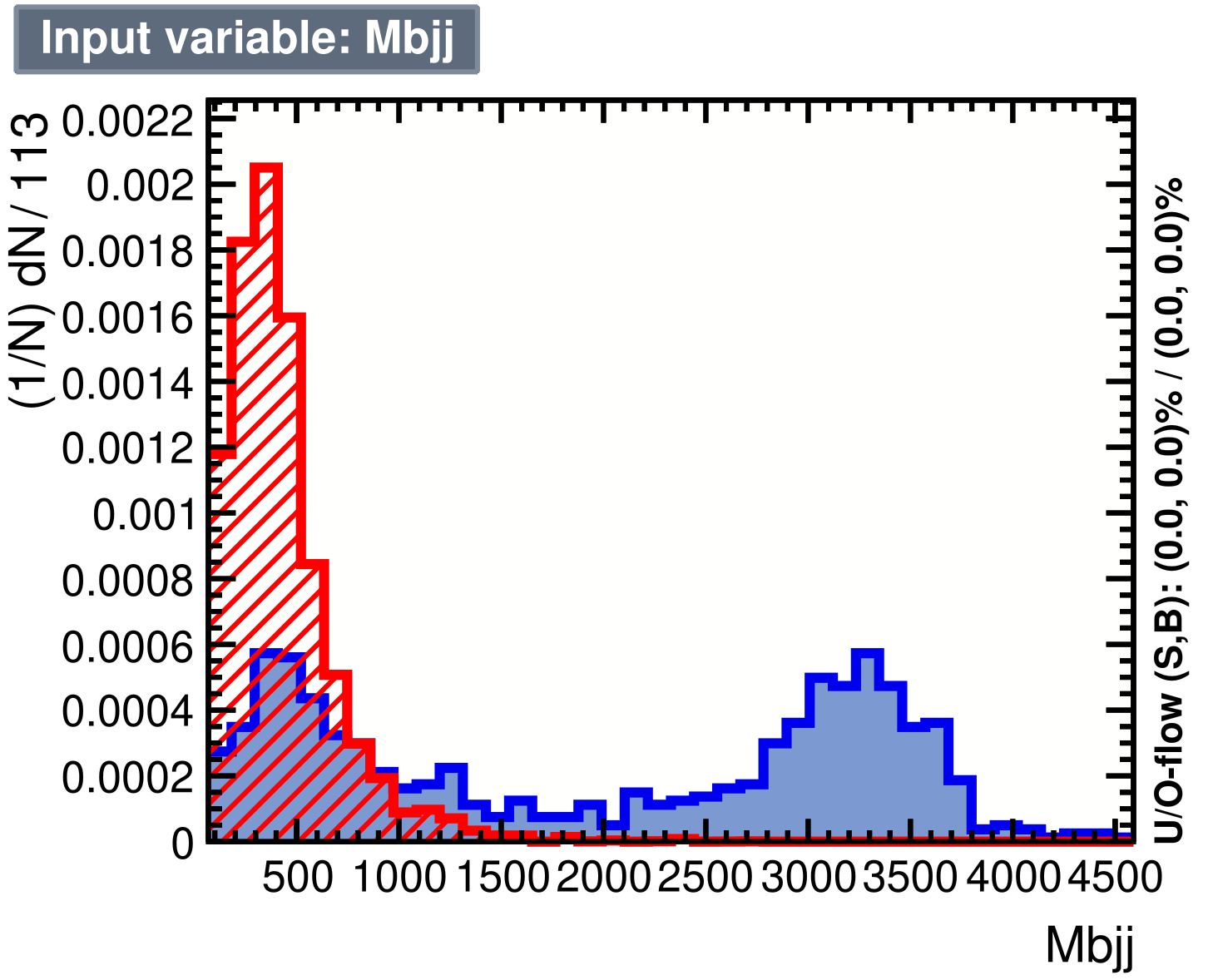}
    \par\smallskip (d)
  \end{minipage}\hfill
  \begin{minipage}[b]{0.28\textwidth}
    \centering
    \includegraphics[width=\textwidth]{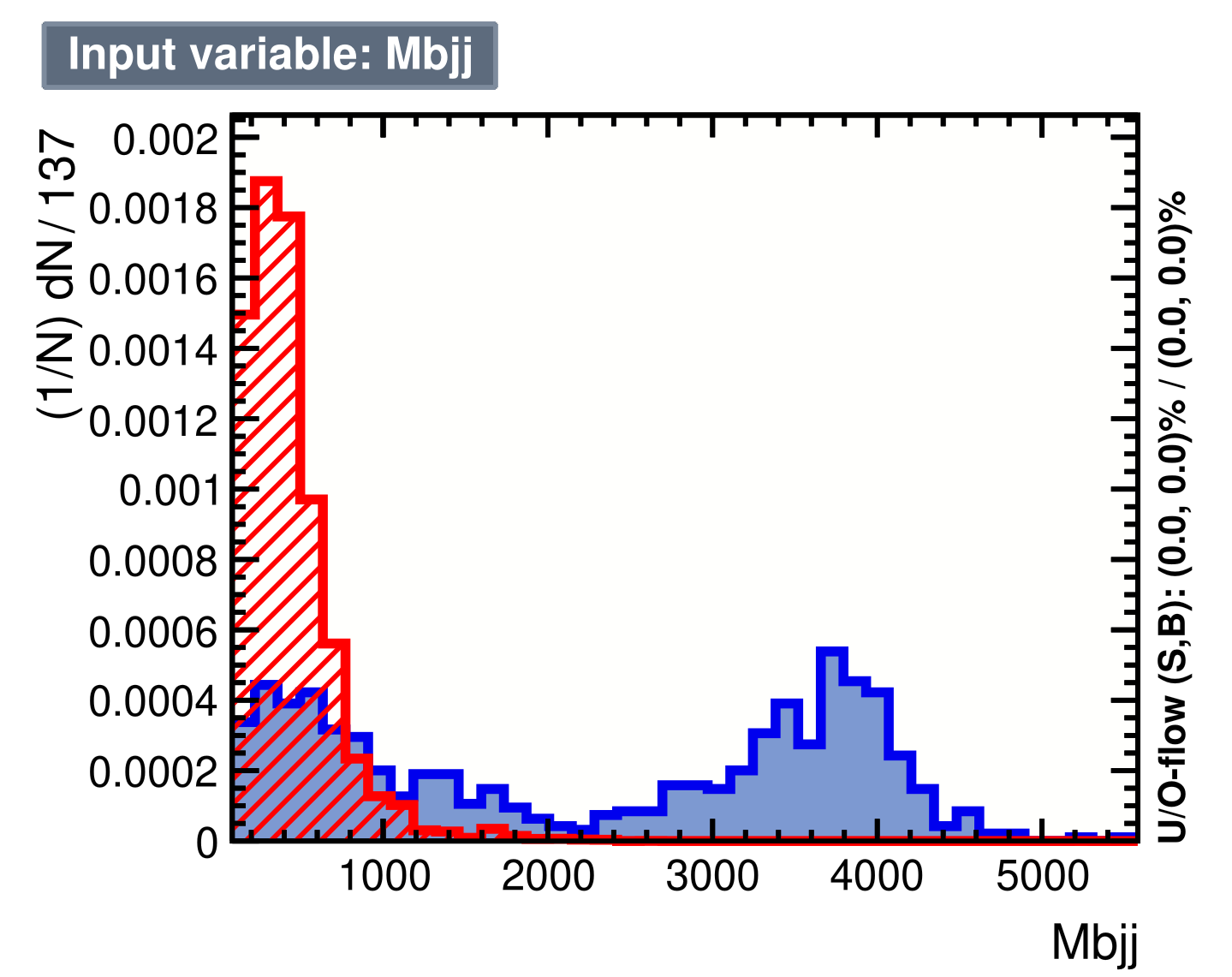}
    \par\smallskip (e)
  \end{minipage}\hfill
  \begin{minipage}[b]{0.28\textwidth}
    \centering
    \includegraphics[width=\textwidth]{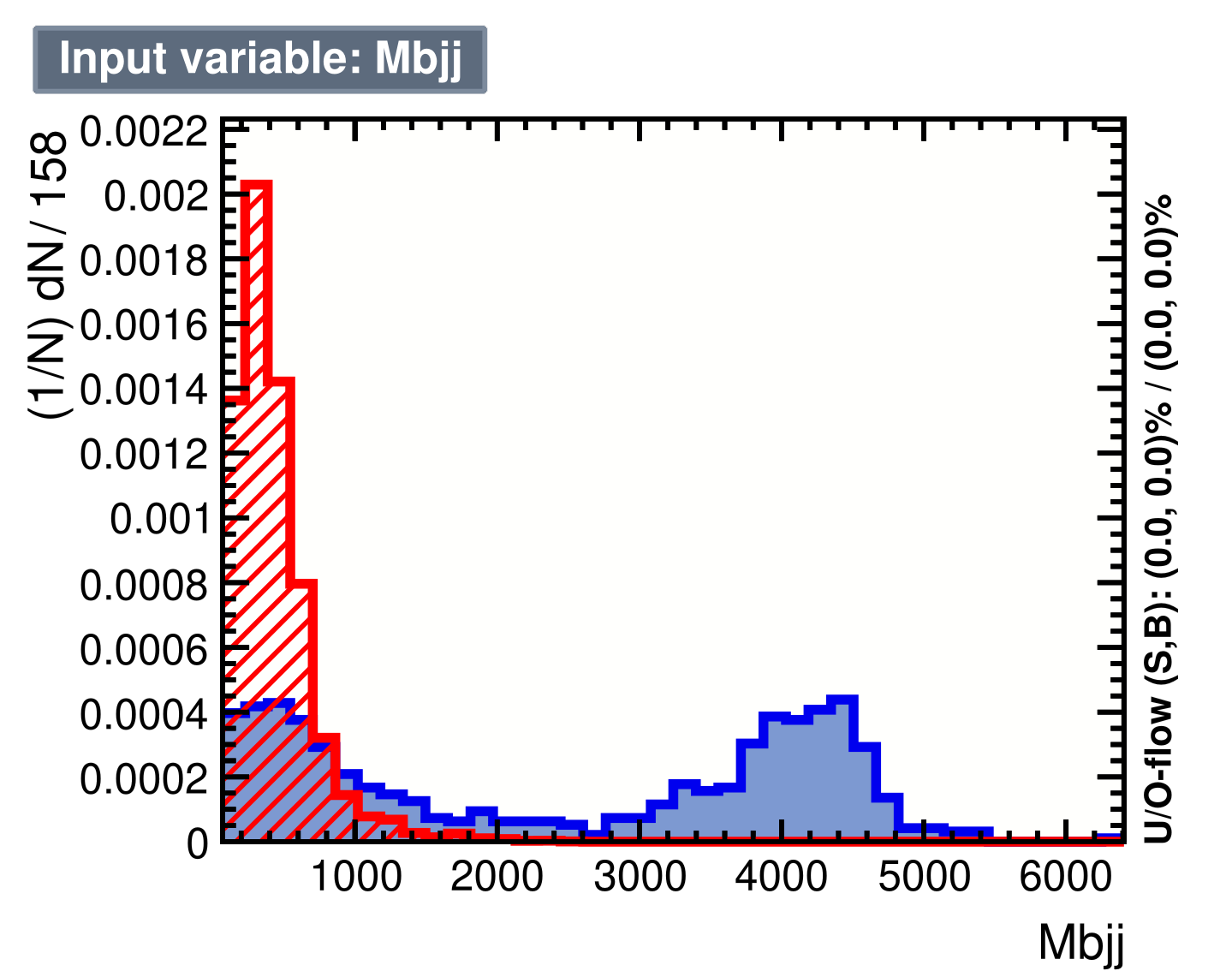}
    \par\smallskip (f)
  \end{minipage}

  \vspace{0.5em} % optional space between rows

  \begin{minipage}[b]{0.28\textwidth}
    \centering
    \includegraphics[width=\textwidth]{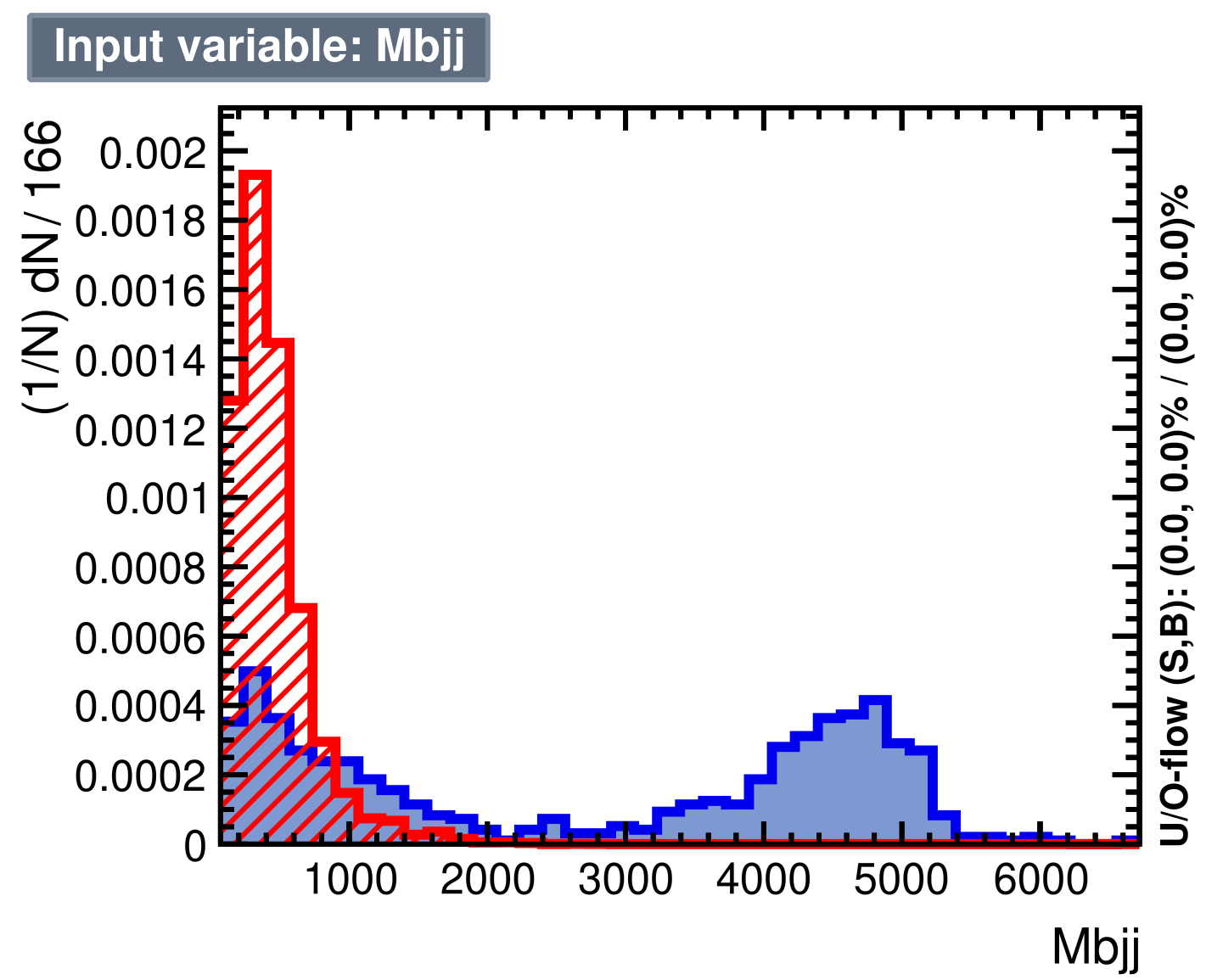}
    \par\smallskip (g)
  \end{minipage}

  \caption{
    The reconstruction of all $m_{T}(2000-5000\, \mathrm{GeV})$ points in the Hadronic Analysis section. It is important to notice that the peaks are close to their required values; nevertheless, a small standard deviation of 100 GeV is also visible.
  }
  \label{fig:6}
\end{figure}

As illustrated in Fig.~\ref{fig:6}, the reconstructed mass distributions exhibit distinct and well-defined peaks corresponding to the benchmark mass points ranging from 2000 GeV to 5000 GeV. The narrowness of these peaks at lower mass points indicates high reconstruction efficiency and excellent energy resolution for the hadronic final states. However, as the mass $m_T$ increases, the decay products become highly boosted. In this regime, the angular separation between the jets from the $W$ boson decay ($\Delta R_{jj}$) decreases, often leading to overlapping jet energy clusters \cite{Marzani:2019hun}. In the present resolved selection, the analysis requires at least four reconstructed jets and two $b$-tags, consistent with the cut-flow tables used throughout the hadronic study.
 %%%%%%%%%%%%%%%%%%%%%%%%%%%%%%%%%%%%

The signal distributions are shown overlaid with the remaining SM background after the baseline selection cuts.  The reconstructed peaks remain identifiable across the scan and stay close to the input benchmark masses, but the efficiency decreases toward the highest masses, where the decay products become increasingly collimated.  The present hadronic mass reconstruction should therefore be interpreted as a resolved-object baseline rather than as a dedicated boosted-$W$ or jet-substructure analysis.  In particular, for $m_T\gtrsim4~\mathrm{TeV}$ the two light jets from the hadronic $W$ can become partially merged, so the reported yields and significances already include the corresponding acceptance loss within the resolved selection.  A future boosted-object optimization could modify the high-mass hadronic performance \cite{Marzani:2019hun}.

  \subsection{The Reconstruction of Mass Points (Leptonic)}
%%%%%%%%%%%%%%%%%%%%%%%%%%%%%%%%%%%%%%%%%%%%%%%%%%%%%%%%%%%%%%%%%%%%%%%%
In the leptonic decay channel, the vector-like $T$ quark mass is reconstructed from a final state consisting of a high-$p_T$ lepton, a $b$-jet, and missing transverse energy ($E_T^{miss}$). The decay chain $T \to Wb \to \ell \nu b$ introduces a significant challenge: the longitudinal momentum of the neutrino ($p_{\nu, z}$) is not directly measurable. To overcome this, we reconstruct the neutrino kinematics by imposing a $W$-boson mass constraint, $M_{\ell\nu} = M_W$, and solving the resulting quadratic equation for $p_{\nu, z}$ \cite{Chatrchyan:2013vlq, Bai:2022up}. In cases where the solution yields complex roots, the real part is typically used, whereas for multiple real solutions, the one that provides the best kinematic fit to the $T$ quark candidate is selected.

As shown in Fig. 9, the invariant mass of the leptonic system, denoted as $M_{bl\nu}$, exhibits clear resonant peaks for the benchmark mass points ranging from 2000 GeV to 5000 GeV. Compared to the hadronic mass points shown in Fig.~\ref{fig:6}, the leptonic peaks appear slightly broader. This broadening is a consequence of the $p_{\nu, z}$ reconstruction ambiguity and the resolution limits of the $E_T^{miss}$ measurement in the presence of the spectator neutrino from the primary production vertex ($\mu p \to T b \nu_\mu$) \cite{Han:2025WbMuP}.

Despite the lower mass resolution, the leptonic channel provides a much cleaner signal-to-background ratio. The high-$p_T$ lepton requirement significantly suppresses the QCD multi-jet backgrounds that dominate the hadronic channel. Furthermore, as the center-of-mass energy increases to 9.16 TeV, the lepton and $b$-jet from the $T$ decay become highly collimated. The isolation requirements for the lepton must be carefully optimized in this boosted regime to avoid accidental overlaps with the $b$-jet energy clusters \cite{Rehermann:2010top, Marzani:2019hun}.

These reconstructed leptonic mass distributions, along with the kinematic variables provided in Table X, serve as critical inputs for the MVA classifiers, ensuring a highly sensitive search across the multi-TeV mass range \cite{Radovic:2018dfi}.
%%%%%%%%%%%%%%%%%%%%%%%%%%%%%%%%%%%%%%%%%%%%%%%%%%%%%%%%%%%%%%%%%%%%%%%%
 \begin{figure}[H]
  \centering
  \begin{minipage}[b]{0.28\textwidth}
  \centering
  \includegraphics[width=\textwidth]{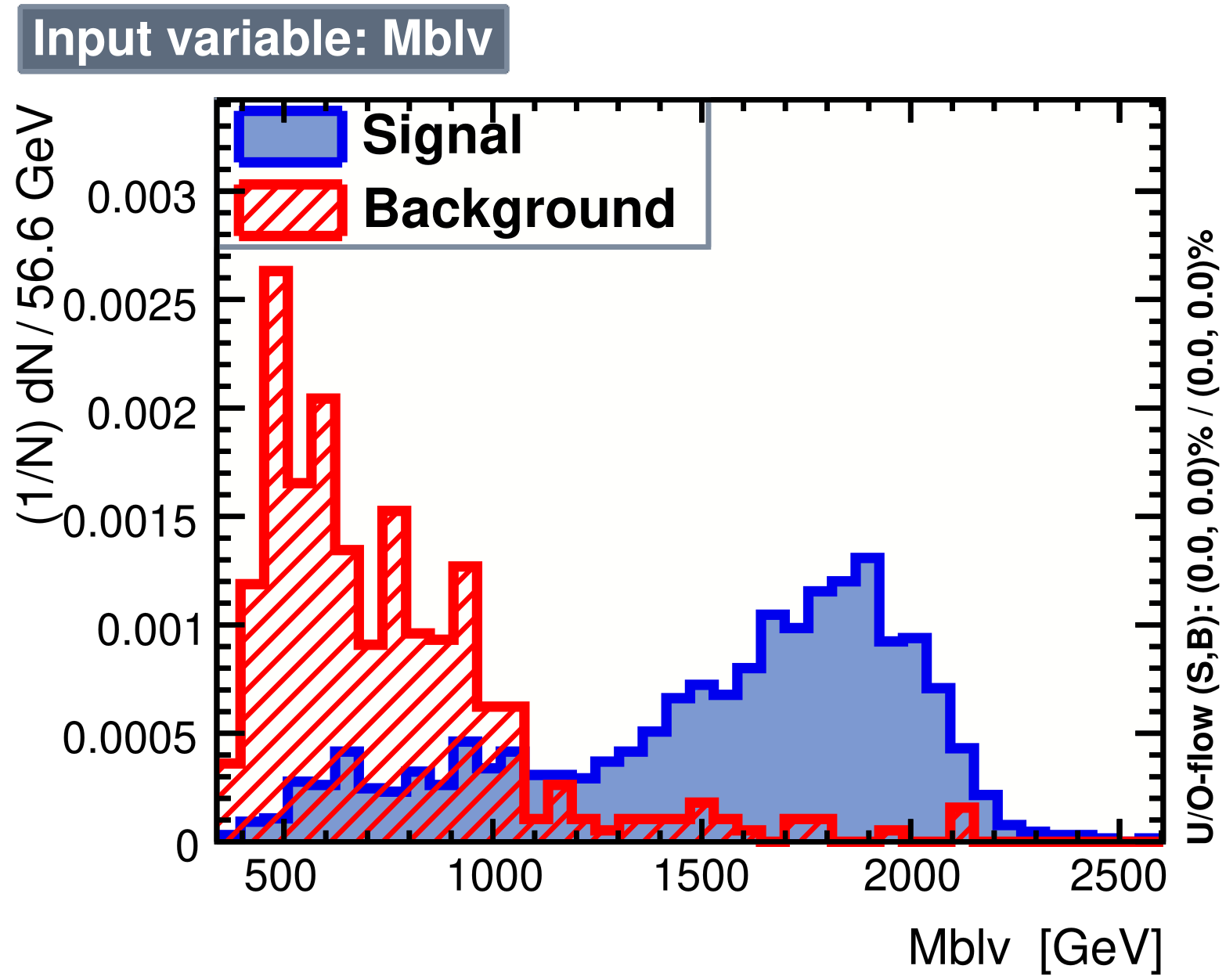}
  \par\small (a)
  \end{minipage}\hfill
  \begin{minipage}[b]{0.28\textwidth}
  \centering
  \includegraphics[width=\textwidth]{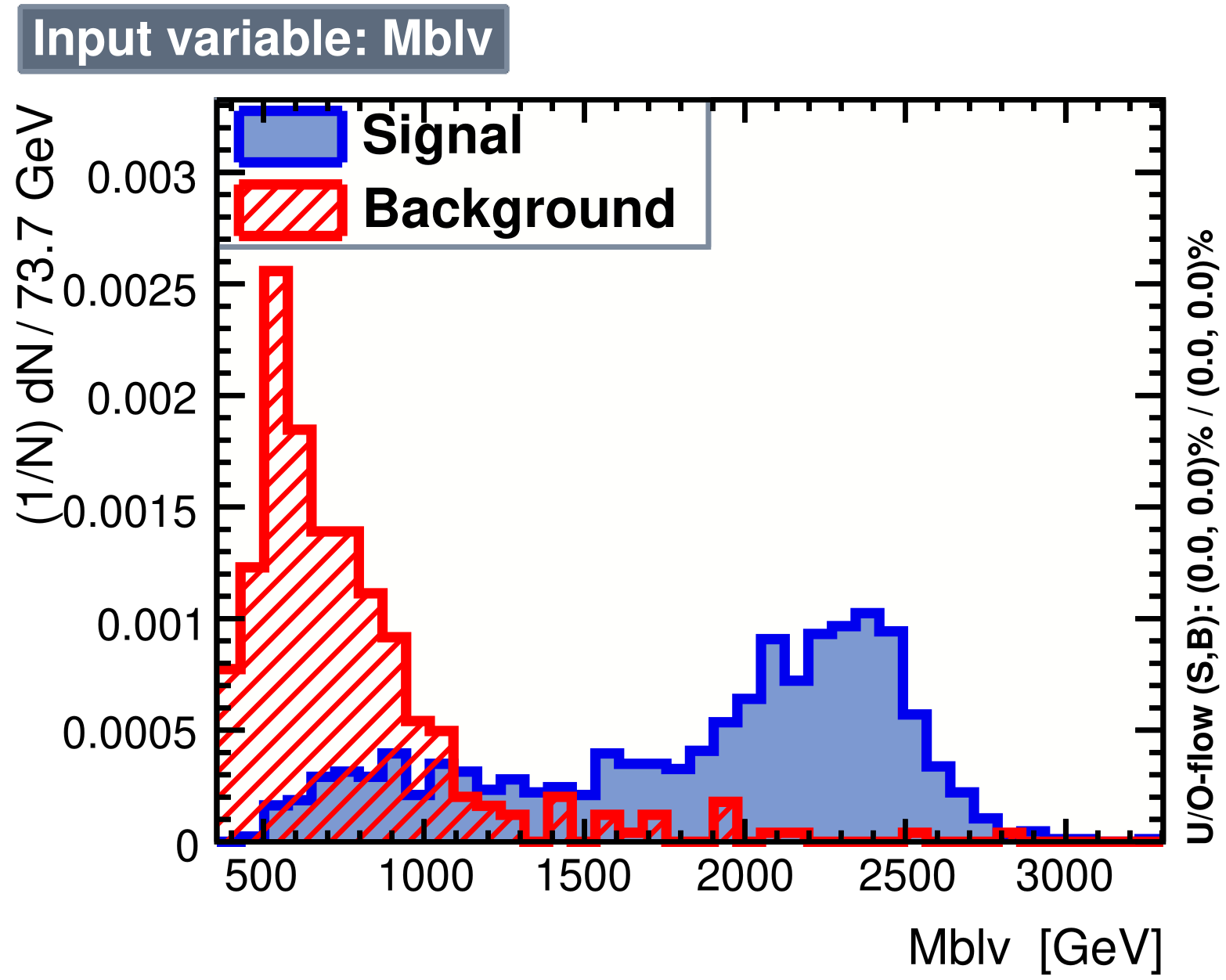}
  \par\small (b)
  \end{minipage}\hfill
  \begin{minipage}[b]{0.28\textwidth}
  \centering
  \includegraphics[width=\textwidth]{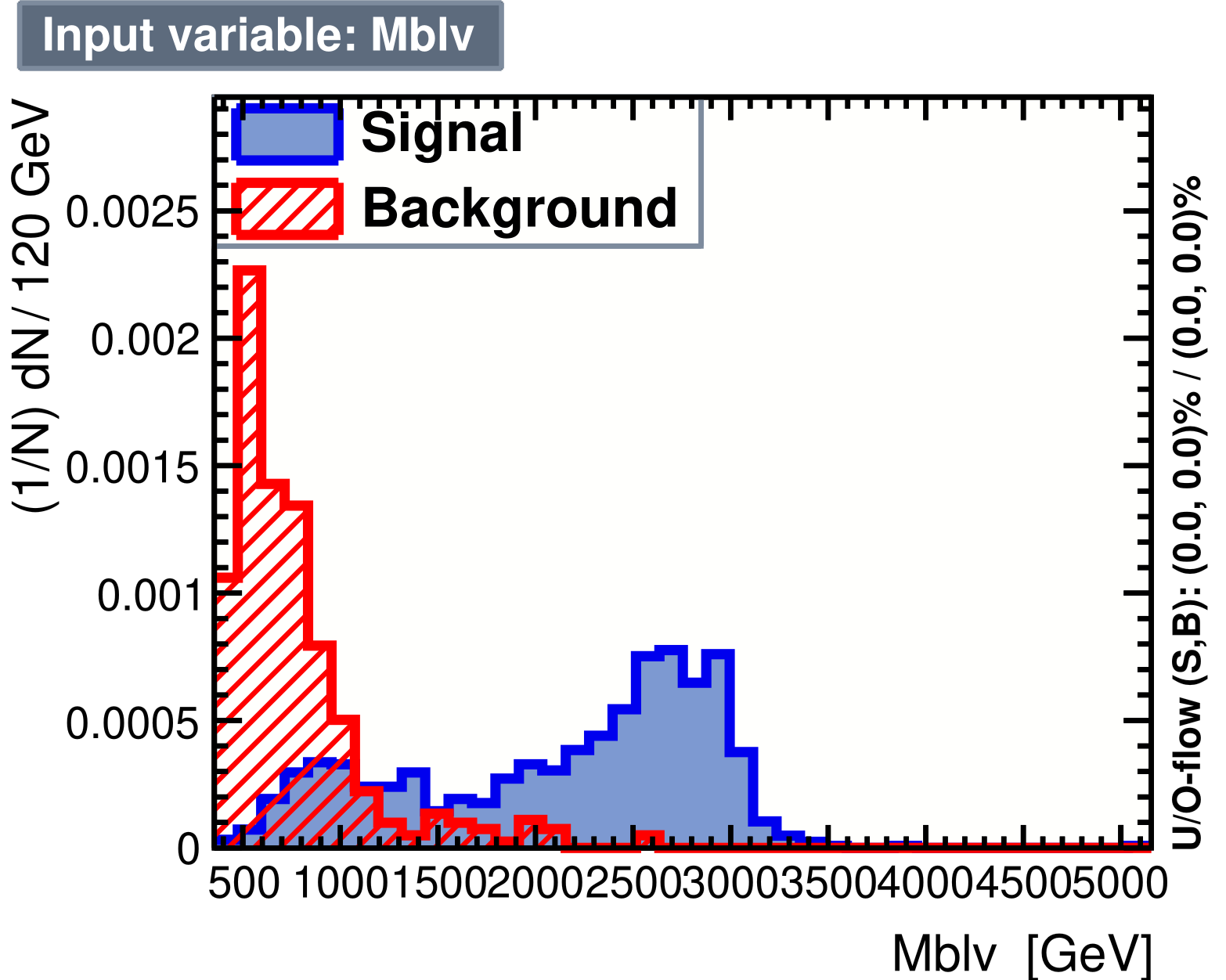}
  \par\small (c)
  \end{minipage}

  \par\smallskip

  \begin{minipage}[b]{0.28\textwidth}
  \centering
  \includegraphics[width=\textwidth]{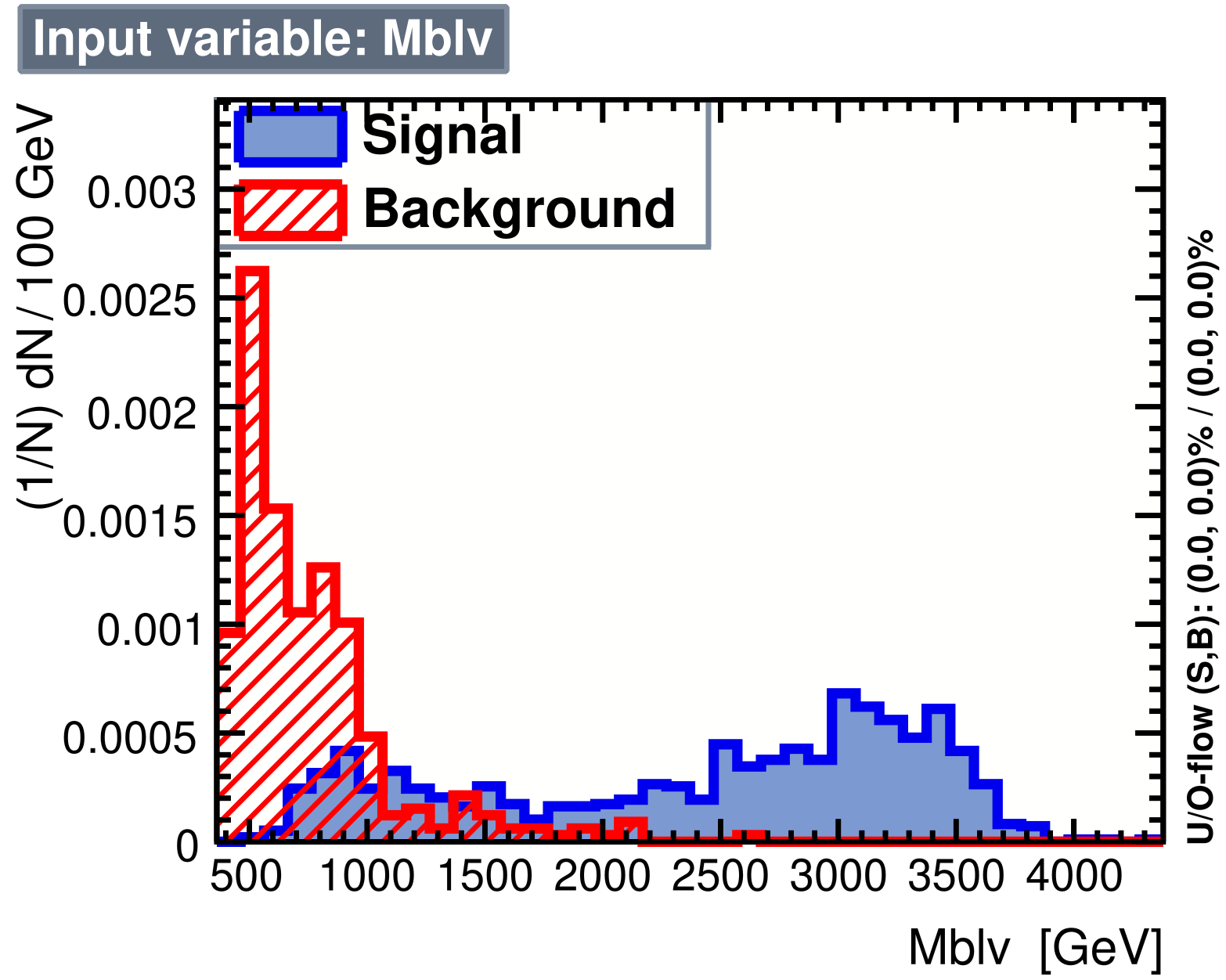}
  \par\small (d)
  \end{minipage}\hfill
  \begin{minipage}[b]{0.28\textwidth}
  \centering
  \includegraphics[width=\textwidth]{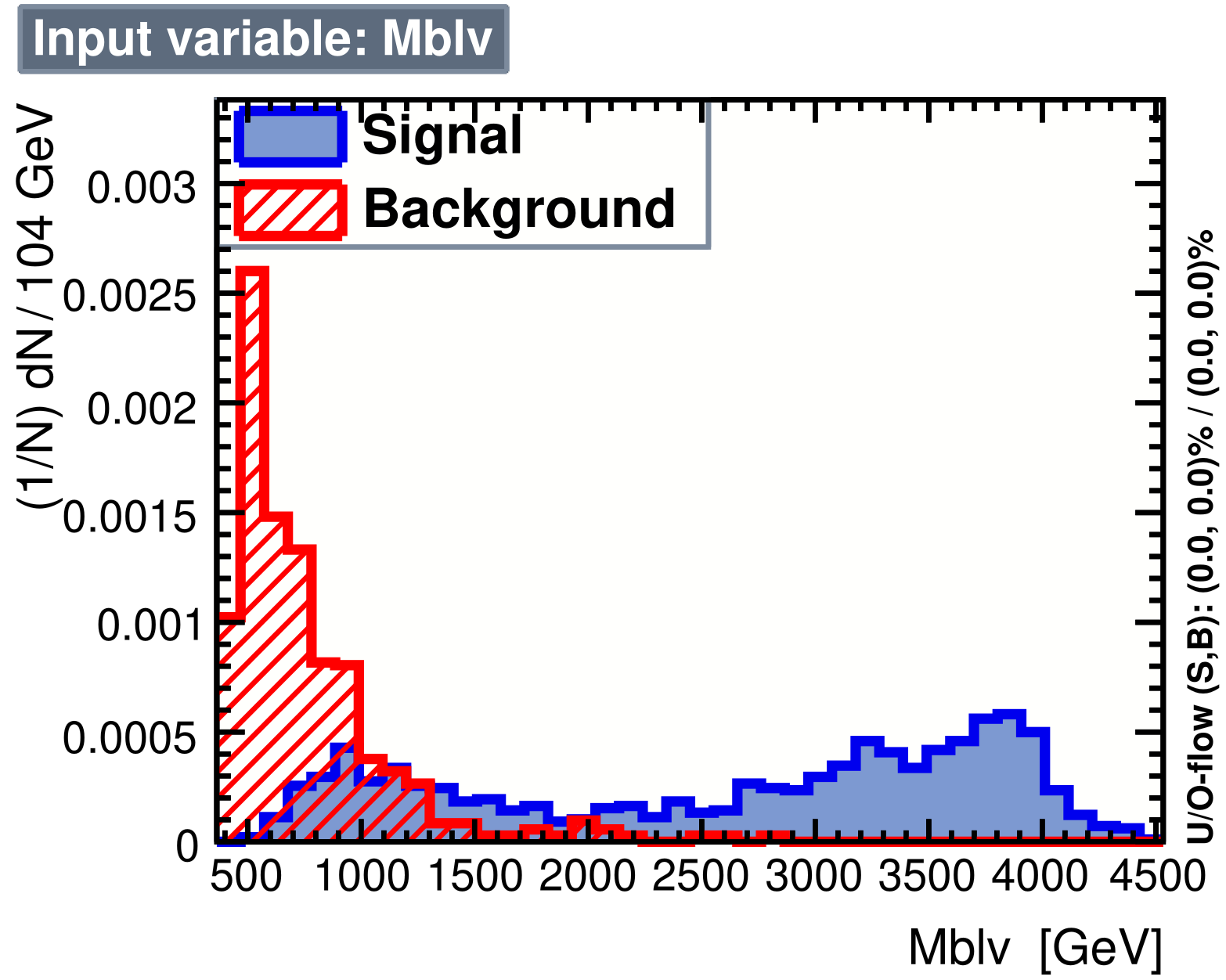}
  \par\small (e)
  \end{minipage}\hfill
  \begin{minipage}[b]{0.28\textwidth}
  \centering
  \includegraphics[width=\textwidth]{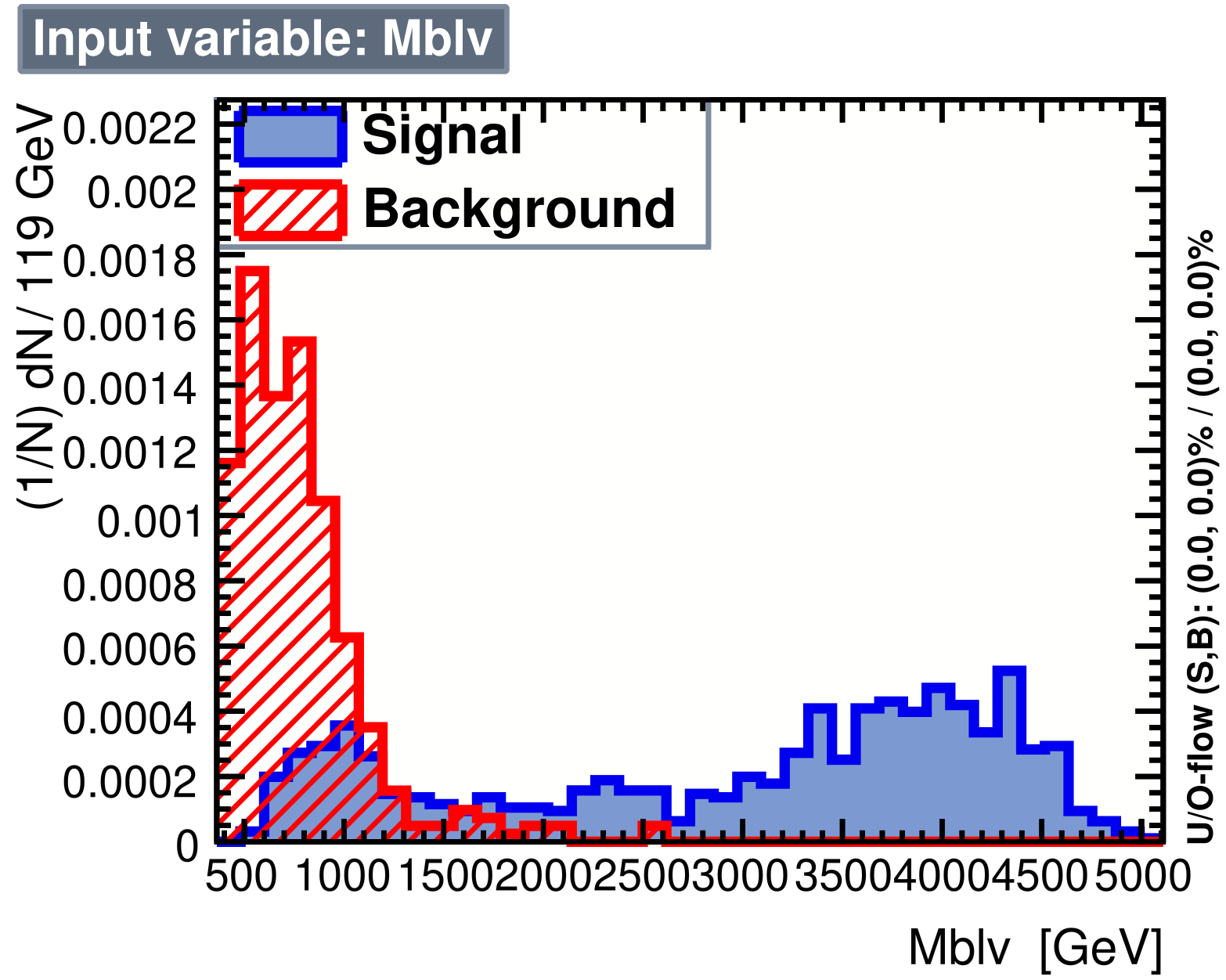}
  \par\small (f)
  \end{minipage}
    \begin{minipage}[b]{0.28\textwidth}
  \centering
  \includegraphics[width=\textwidth]{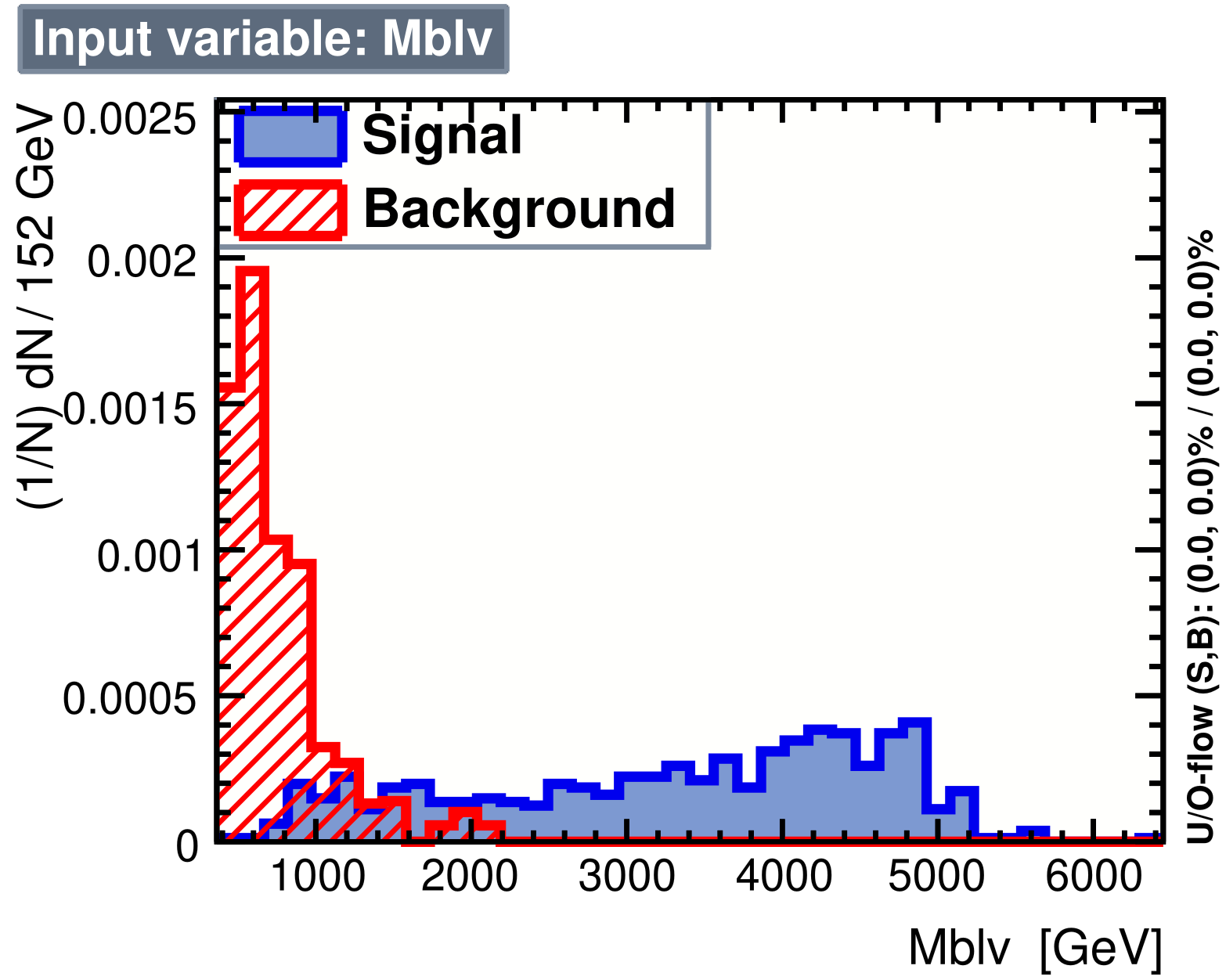}
  \par\small (g)
  \end{minipage}\hfill

  \caption{Reconstructed $m_T$ distributions for the leptonic analysis across the $2000$--$5000$ GeV benchmark points. The peaks remain close to the input masses, and the widths are generally smaller than in the hadronic channel because of the cleaner final-state signature.}
  \label{fig:7}
  \end{figure}

  \subsection{The Statistical Significance of Hadronic and Leptonic Processes}

The task of evaluating the statistical significance ($\sigma$) for the vector-like singlet top ($T$) in the fully hadronic and leptonic final states presents a complex challenge due to the overwhelming backgrounds. To optimize the sensitivity ($Z$), the optimal cuts were carefully chosen and applied to the distributions of four multivariate classifiers (BDT, BDTG, MLP, Likelihood) after successful training in both analysis sections. The cuts were kept constant while calculating the significance at specific mass points and center-of-mass energies across all three luminosities. Finally, the significance $Z$ was calculated using:

\begin{equation}
    Z = \frac{S}{\sqrt{S + B}}
\end{equation}

To quantify the effect of background-normalization systematics, we also use the Cowan--Cranmer--Gross--Vitells profile-likelihood Asimov significance \cite{Cowan:2010js}.  For a fractional background uncertainty $\epsilon_B=20\%$, we take $\sigma_B=\epsilon_B B$ and compute
\begin{equation}
Z_A =
\left\{
2\left[
(S+B)\ln\!\left(\frac{(S+B)(B+\sigma_B^2)}{B^2+(S+B)\sigma_B^2}\right)
- \frac{B^2}{\sigma_B^2}\ln\!\left(1+\frac{\sigma_B^2 S}{B(B+\sigma_B^2)}\right)
\right]
\right\}^{1/2}.
\label{eq:asimov}
\end{equation}
The benchmark comparison tables in Sec.~\ref{sec:cutbased_baseline} list the corresponding post-selection $S$ and $B$ values explicitly for the literature baseline and the present MVA working point.  The large nominal significances quoted elsewhere in the manuscript should therefore be read as fast-simulation projections under the stated generator, showering, detector, object-selection, and MVA assumptions.  To expose the impact of background-normalization systematics explicitly, we also report $Z_A$ with the assumed $20\%$ normalization uncertainty.

For luminosity consistency in the full scan, the classifier and working point selected at $3000~\mathrm{fb}^{-1}$ are held fixed.  Where the raw scan yields are not tabulated explicitly, the effective $S/B$ is algebraically reconstructed from the displayed $(Z,Z_A)$ pair at $3000~\mathrm{fb}^{-1}$ using Eqs.~(\ref{eq:asimov}) and $Z=S/\sqrt{S+B}$.  The resulting $S$ and $B$ are then scaled linearly with luminosity, and both $Z$ and $Z_A$ are recomputed.  Thus the $100~\mathrm{fb}^{-1}$ entries below are not independently re-optimized and obey the stated fixed-working-point prescription.

\begin{table}[htbp]
\centering
\caption{Comparative summary of the statistical performance at $\mathcal{L} = 100$ fb$^{-1}$. The values are presented as \textbf{Gaussian Significance ($Z$)} and \textbf{Asimov Significance ($Z_A$)}.  To maintain one luminosity-independent signal region, the classifier and working point are those of Table~\ref{tab:best_vs_best_dual}; the corresponding signal and background yields are scaled by $100/3000$ before recomputing both significances.}
\label{tab:best_vs_best_100fb}
\small
\renewcommand{\arraystretch}{1.15}
\setlength{\tabcolsep}{4pt}
\resizebox{\textwidth}{!}{%
\begin{tabular}{|l|c|c|c|c|c|c|c|c|c|c|c|c|}
\hline
\textbf{Mass} & \multicolumn{4}{c|}{\textbf{$\sqrt{s} = 5.29$ TeV}} & \multicolumn{4}{c|}{\textbf{$\sqrt{s} = 6.48$ TeV}} & \multicolumn{4}{c|}{\textbf{$\sqrt{s} = 9.16$ TeV}} \\
\hline
($M_T$) & \multicolumn{2}{c|}{\textbf{Hadronic}} & \multicolumn{2}{c|}{\textbf{Leptonic}} & \multicolumn{2}{c|}{\textbf{Hadronic}} & \multicolumn{2}{c|}{\textbf{Leptonic}} & \multicolumn{2}{c|}{\textbf{Hadronic}} & \multicolumn{2}{c|}{\textbf{Leptonic}} \\
\hline
& \textbf{$Z$} & \textbf{$Z_A$} & \textbf{$Z$} & \textbf{$Z_A$} & \textbf{$Z$} & \textbf{$Z_A$} & \textbf{$Z$} & \textbf{$Z_A$} & \textbf{$Z$} & \textbf{$Z_A$} & \textbf{$Z$} & \textbf{$Z_A$} \\
\hline
2000 GeV &
16.62 (BDT) & 21.83 (BDT) & 6.41 (BDTG) & 6.10 (BDTG) &
23.67 (BDTG) & 5.88 (BDTG) & 12.16 (BDTG) & 14.12 (BDTG) &
61.40 (MLP) & 12.63 (MLP) & 26.73 (BDT) & 18.45 (BDT) \\
\hline
2500 GeV &
9.57 (BDT) & 9.55 (BDT) & 3.07 (Likelihood) & 4.62 (Likelihood) &
16.54 (BDT) & 5.88 (BDT) & 8.59 (BDTG) & 9.25 (BDTG) &
58.52 (MLP) & 23.83 (MLP) & 24.59 (BDT) & 20.53 (BDT) \\
\hline
3000 GeV &
3.07 (BDTG) & 1.86 (BDTG) & 1.47 (BDTG) & 1.61 (BDTG) &
8.25 (MLP) & 1.76 (MLP) & 5.57 (BDT) & 6.72 (BDT) &
49.94 (MLP) & 38.70 (MLP) & 20.79 (MLP) & 18.59 (MLP) \\
\hline
3500 GeV &
0.39 (BDTG) & 0.19 (BDTG) & 0.68 (BDT) & 0.91 (BDT) &
3.06 (BDTG) & 0.63 (BDTG) & 2.69 (BDTG) & 3.84 (BDTG) &
38.90 (BDT) & 22.78 (BDT) & 16.29 (BDT) & 17.49 (BDT) \\
\hline
4000 GeV &
0.05 (MLP) & 0.04 (MLP) & 0.21 (BDT) & 0.24 (BDT) &
2.44 (MLP) & 1.77 (MLP) & 2.55 (BDT) & 2.61 (BDT) &
30.20 (BDT) & 24.39 (BDT) & 12.43 (BDT) & 19.44 (BDT) \\
\hline
4500 GeV &
-- & -- & -- & -- &
0.13 (MLP) & 0.03 (MLP) & 0.24 (BDTG) & 0.24 (BDTG) &
20.28 (BDTG) & 13.22 (BDTG) & 8.49 (BDT) & 12.59 (BDT) \\
\hline
5000 GeV &
-- & -- & -- & -- &
0.03 (BDTG) & 0.02 (BDTG) & 0.03 (MLP) & 0.03 (MLP) &
12.44 (BDTG) & 7.85 (BDTG) & 5.35 (BDT) & 8.40 (BDT) \\
\hline
\end{tabular}
}
\end{table}

\begin{table}[htbp]
  \centering
  \caption{Comparative summary of the maximum statistical performance at $\mathcal{L} = 3000$ fb$^{-1}$. The values are presented as \textbf{Gaussian Significance ($Z$) / Asimov Significance ($Z_A$)}. The best performing algorithm for each mass point is indicated in parentheses. The comparison between $Z$ and $Z_A$ illustrates the impact of the assumed 20$\%$ background-normalization uncertainty.}
  \label{tab:best_vs_best_dual}
  \small
  \renewcommand{\arraystretch}{1.15}
  \setlength{\tabcolsep}{4pt}
  \begin{adjustbox}{max width=\textwidth}
  \begin{tabular}{|l|c|c|c|c|c|c|c|c|c|c|c|c|}
  \hline
  \textbf{Mass} &
  \multicolumn{4}{c|}{\textbf{$\sqrt{s} = 5.29$ TeV}} &
  \multicolumn{4}{c|}{\textbf{$\sqrt{s} = 6.48$ TeV}} &
  \multicolumn{4}{c|}{\textbf{$\sqrt{s} = 9.16$ TeV}} \\
  \hline
  ($M_T$) &
  \multicolumn{2}{c|}{\textbf{Hadronic}} & \multicolumn{2}{c|}{\textbf{Leptonic}} &
  \multicolumn{2}{c|}{\textbf{Hadronic}} & \multicolumn{2}{c|}{\textbf{Leptonic}} &
  \multicolumn{2}{c|}{\textbf{Hadronic}} & \multicolumn{2}{c|}{\textbf{Leptonic}} \\
  \hline
  & \textbf{$Z$} & \textbf{$Z_A$} & \textbf{$Z$} & \textbf{$Z_A$} &
    \textbf{$Z$} & \textbf{$Z_A$} & \textbf{$Z$} & \textbf{$Z_A$} &
    \textbf{$Z$} & \textbf{$Z_A$} & \textbf{$Z$} & \textbf{$Z_A$} \\
  \hline
  2000 GeV &
  91.04 (BDT) & 29.62 (BDT) & 35.11 (BDTG) & 8.07 (BDTG) &
  129.64 (BDTG) & 5.98 (BDTG) & 66.63 (BDTG) & 18.72 (BDTG) &
  336.28 (MLP) & 12.74 (MLP) & 146.41 (BDT) & 20.10 (BDT) \\
  \hline
  2500 GeV &
  52.42 (BDT) & 12.22 (BDT) & 16.83 (Likelihood) & 11.09 (Likelihood) &
  90.58 (BDT) & 6.09 (BDT) & 47.03 (BDTG) & 12.43 (BDTG) &
  320.52 (MLP) & 24.48 (MLP) & 134.70 (BDT) & 23.19 (BDT) \\
  \hline
  3000 GeV &
  16.84 (BDTG) & 2.19 (BDTG) & 8.03 (BDTG) & 3.44 (BDTG) &
  45.20 (MLP) & 1.79 (MLP) & 30.49 (BDT) & 10.52 (BDT) &
  273.51 (MLP) & 42.23 (MLP) & 113.85 (MLP) & 21.51 (MLP) \\
  \hline
  3500 GeV &
  2.16 (BDTG) & 0.21 (BDTG) & 3.75 (BDT) & 3.75 (BDT) &
  16.76 (BDTG) & 0.64 (BDTG) & 14.74 (BDTG) & 9.22 (BDTG) &
  213.04 (BDT) & 24.12 (BDT) & 89.20 (BDT) & 21.75 (BDT) \\
  \hline
  4000 GeV &
  0.29 (MLP) & 0.06 (MLP) & 1.13 (BDT) & 1.13 (BDT) &
  13.39 (MLP) & 2.26 (MLP) & 13.98 (BDT) & 4.31 (BDT) &
  165.43 (BDT) & 27.15 (BDT) & 68.07 (BDT) & 30.76 (BDT) \\
  \hline
  4500 GeV &
  -- & -- & -- & -- &
  0.72 (MLP) & 0.03 (MLP) & 1.29 (BDTG) & 0.67 (BDTG) &
  111.09 (BDTG) & 14.43 (BDTG) & 46.49 (BDT) & 20.71 (BDT) \\
  \hline
  5000 GeV &
  -- & -- & -- & -- &
  0.19 (BDTG) & 0.02 (BDTG) & 0.16 (MLP) & 0.08 (MLP) &
  68.16 (BDTG) & 8.70 (BDTG) & 29.31 (BDT) & 16.70 (BDT) \\
  \hline
  \end{tabular}
  \end{adjustbox}
  \end{table}

\begin{figure}[H]
\centering
\begin{subfigure}[b]{0.50\textwidth}
\centering
\includegraphics[width=7cm, height=6cm]{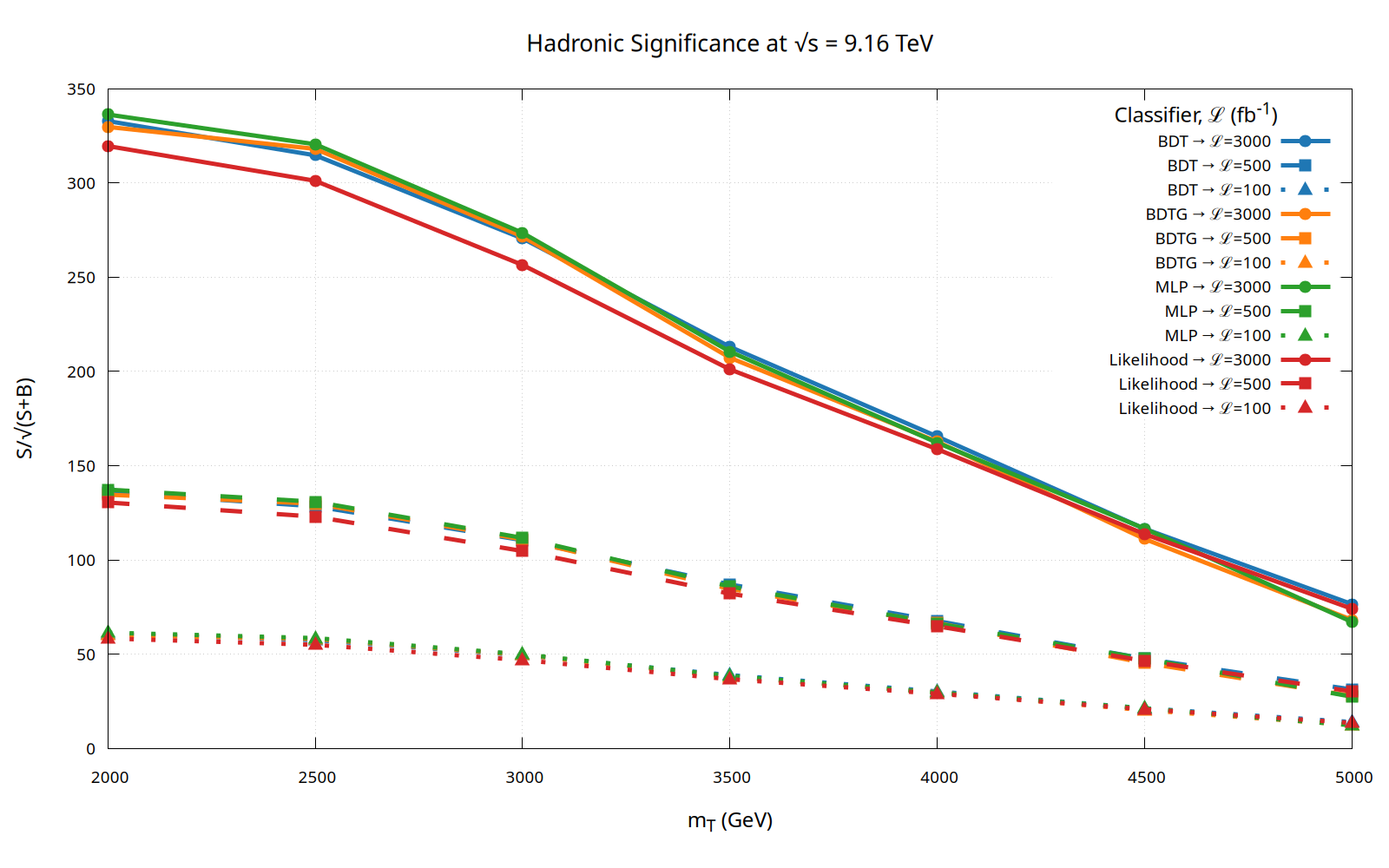}
\caption{Hadronic}
\end{subfigure}\hfill
\begin{subfigure}[b]{0.50\textwidth}
\centering
\includegraphics[width=7cm, height=6cm]{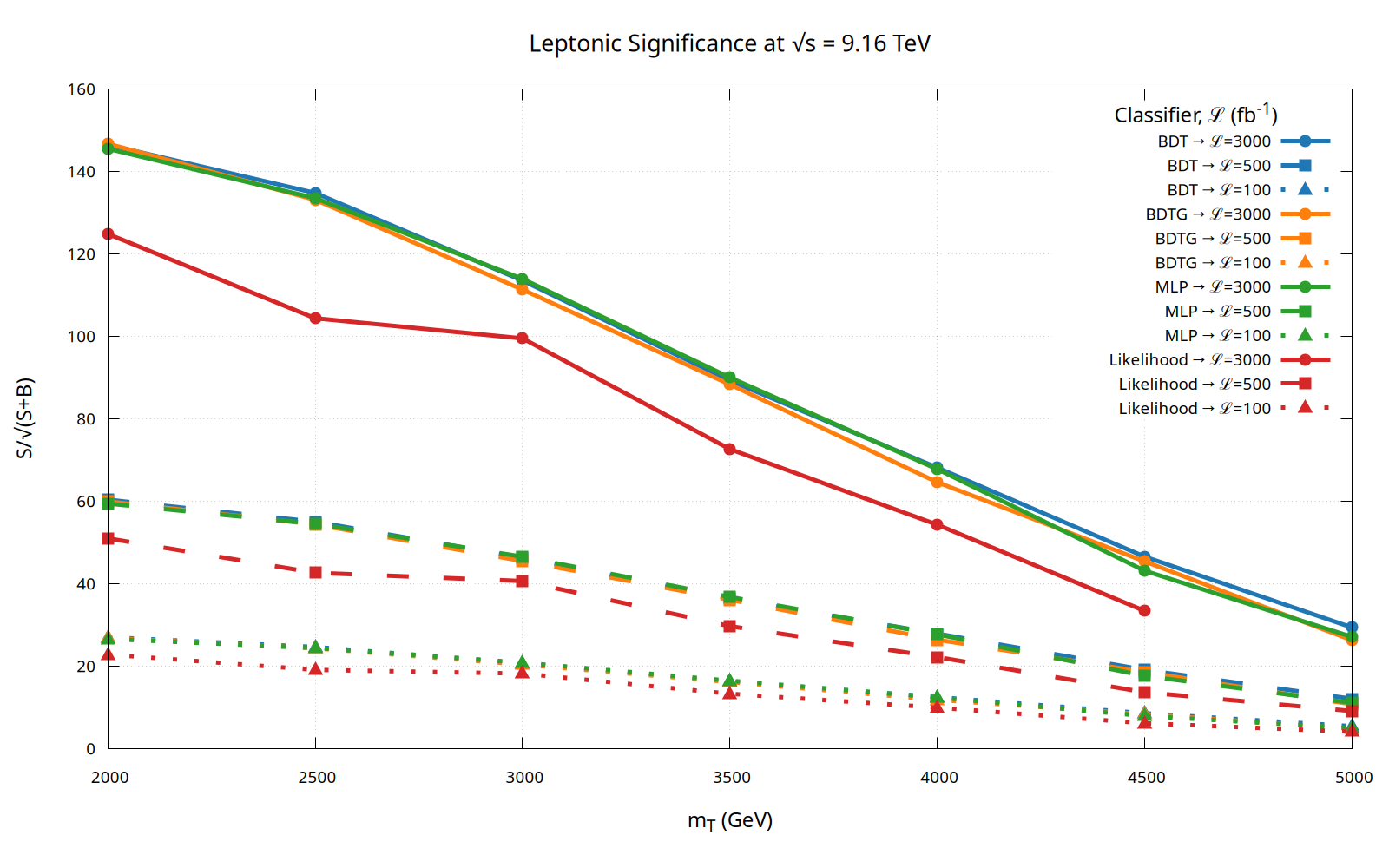}
\caption{Leptonic}
\end{subfigure}
\caption{Hadronic and leptonic signal significances at $\sqrt{s}=9.16~\mathrm{TeV}$ for all three luminosities.}
\label{fig:signif-had}
\end{figure}
 \subsection{Interpretation of Results and Discussion}
\begin{figure}[H]
    \centering
    \begin{subfigure}[b]{0.48\linewidth}
      \centering
      \includegraphics[width=\linewidth]{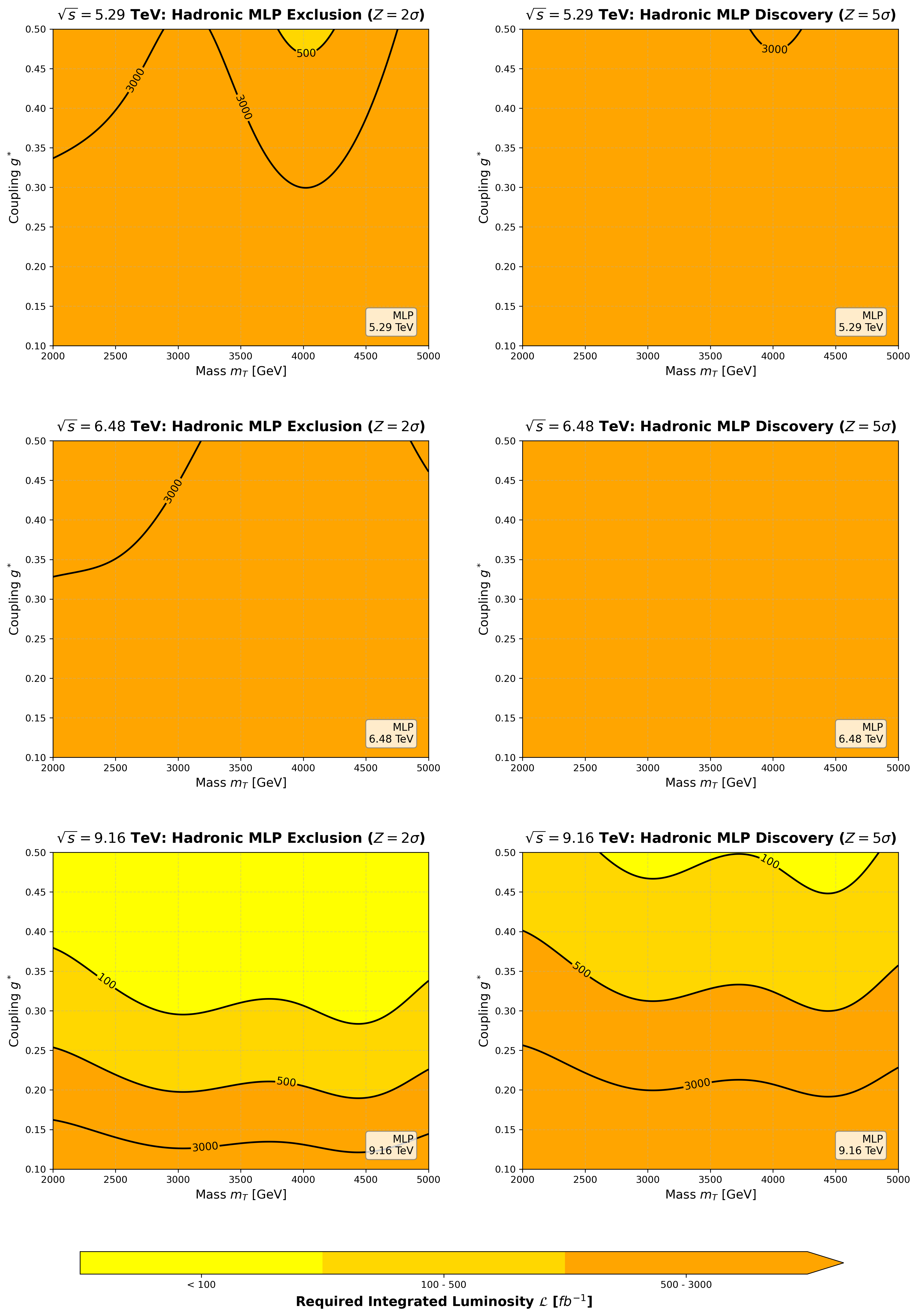}
      \caption{Sensitivity projections in the hadronic channel using the MLP classifier for $\mu^- p \to \nu_\mu T \bar{b} \to \nu_\mu (Wb)\bar{b} \to \nu_\mu (jjb)\bar{b}$}
      \label{fig:rd-hadronic}
    \end{subfigure}\hfill%
    \begin{subfigure}[b]{0.48\linewidth}
      \centering
      \includegraphics[width=\linewidth]{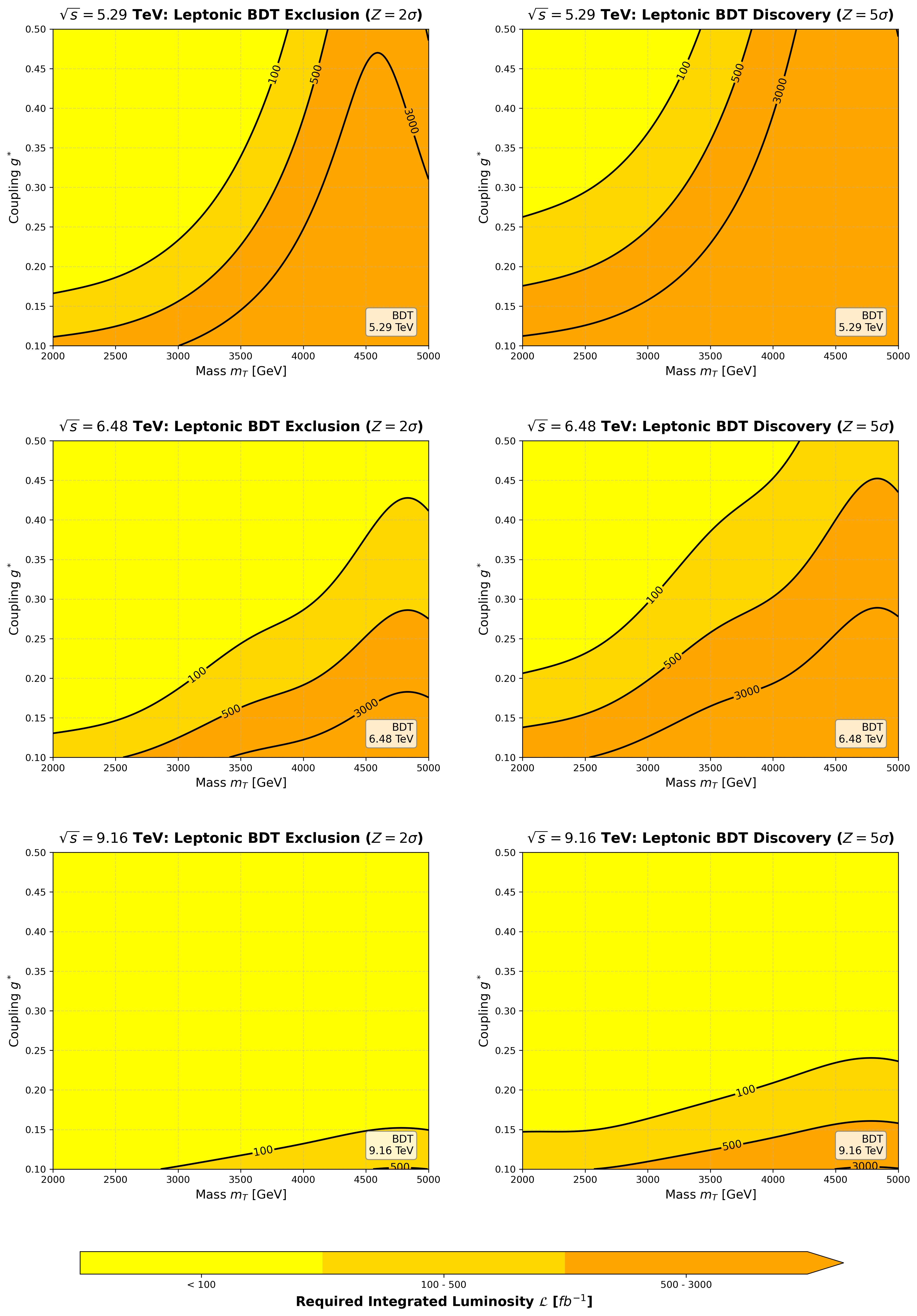}
      \caption{Sensitivity projections in the leptonic channel using the BDT classifier for $\mu^- p \to \nu_\mu T \bar{b} \to \nu_\mu (Wb)\bar{b} \to \nu_\mu (\ell \nu b)\bar{b}$}
      \label{fig:rd-leptonic}
    \end{subfigure}
   \caption{Projected exclusion and discovery contours in the $g^{*}$--$m_T$ plane.  Panel (a) shows the hadronic analysis with the MLP classifier and panel (b) shows the leptonic analysis with the BDT classifier, both evaluated across the benchmark center-of-mass energies and luminosities used in this study.}
    \label{fig:side-by-side-results}
  \end{figure}

The statistical landscape of Vector-Like Top quark ($T$) searches at the proposed $\mu p$ collider is governed by a non-trivial interplay between integrated luminosity, center-of-mass energy, and the intrinsic characteristics of the hadronic and leptonic decay channels. By tracking the evolution of sensitivity from the early data regime in Table~XI($\mathcal{L}=100~\mathrm{fb}^{-1}$) to the high-luminosity frontier summarized in Table~XII($\mathcal{L}=3000~\mathrm{fb}^{-1}$), several robust physical patterns emerge. While the Gaussian significance follows the expected $Z\propto\sqrt{\mathcal{L}}$ scaling, the comparison with the Asimov significance $Z_A$ reveals a qualitatively different behavior at high statistics. In particular, channels with large raw yields eventually encounter a systematics-dominated regime, where further luminosity accumulation no longer translates into proportional gains in projected sensitivity. This effect is clearly visible at $\sqrt{s}=9.16~\mathrm{TeV}$ in the hadronic channel for $m_T=3000~\mathrm{GeV}$, where the Gaussian significance reaches $273.51\sigma$ while the corresponding $Z_A$ saturates at $42.23\sigma$, indicating that a $20\%$ background systematic uncertainty sets the ultimate sensitivity ceiling.\\
These trends are visualized in the luminosity contour maps shown in Fig. \ref{fig:side-by-side-results}, which display the required integrated luminosity for exclusion ($Z=2\sigma$) and discovery ($Z=5\sigma$) in the $(m_T,g^*)$ plane. The hadronic-channel MLP sensitivity map demonstrates that, despite its large signal yield driven by the dominant branching fraction, the presence of sizable QCD backgrounds pushes the required luminosity contours outward, particularly once systematic uncertainties are accounted for. In contrast, the leptonic-channel BDT sensitivity map exhibits substantially lower luminosity requirements across a wide region of parameter space, reflecting the cleaner experimental signature associated with an isolated high-$p_T$ lepton and missing transverse energy. As a result, the leptonic channel retains projected sensitivity even in regions where signal event counts become sparse.\\

Figure~\ref{fig:side-by-side-results} summarizes the integrated-luminosity contours in the $(g^{*},m_T)$ plane using the Asimov significance with a $20\%$ background-normalization uncertainty.  Panel (a) shows the fully hadronic MLP analysis, where the larger branching fraction gives the strongest sensitivity at intermediate masses but the required luminosity increases once residual QCD backgrounds and resolved-reconstruction losses are included.  Panel (b) shows the leptonic BDT analysis, whose cleaner final state sustains sensitivity to larger masses and lower couplings within the present assumptions.  The contours delineate the regions accessible at the $2\sigma$ and $5\sigma$ levels.

A recurring feature across all luminosity benchmarks is the crossover between hadronic and leptonic dominance as a function of $m_T$. At intermediate masses, typically between $2000$ and $3000~\mathrm{GeV}$, the hadronic channel provides the strongest sensitivity owing to its superior production rate, which compensates for the larger backgrounds. This behavior is especially pronounced at $\sqrt{s}=9.16~\mathrm{TeV}$, where the hadronic analysis yields the highest nominal significances of the entire study. As the mass approaches the kinematic limit, however, the leptonic channel becomes increasingly competitive and ultimately dominant. Its stronger background suppression leads to a higher signal-to-background ratio, making it more resilient in the low-statistics regime that characterizes the extreme mass frontier. Consequently, for $m_T$ values near $5000~\mathrm{GeV}$, the leptonic channel consistently outperforms the hadronic channel across all luminosity scenarios.
The dependence on center-of-mass energy further highlights the critical role of the $\sqrt{s}=9.16~\mathrm{TeV}$ configuration. At $\sqrt{s}=5.29~\mathrm{TeV}$ the sensitivity rapidly deteriorates beyond $m_T\sim3000~\mathrm{GeV}$ as the production cross section approaches the kinematic threshold. By contrast, the $9.16~\mathrm{TeV}$ benchmark sustains projected significance above the discovery threshold over a much broader mass range under the stated assumptions, in many cases even with early data samples. This indicates that, for heavy Vector-Like Quark searches in the present setup, increasing the collision energy is substantially more effective than luminosity accumulation alone. Algorithmically, a complementary pattern emerges: Boosted Decision Trees dominate the leptonic analysis by efficiently exploiting a relatively low-dimensional feature space, while Multi-Layer Perceptrons become increasingly advantageous in the hadronic channel, where their ability to model non-linear correlations among multi-jet observables enhances background rejection.

At the highest-energy and highest-luminosity benchmark of $\sqrt{s}=9.16~\mathrm{TeV}$ with $\mathcal{L}=3000~\mathrm{fb}^{-1}$, the sensitivity maps allow a concise quantitative summary of the achievable reach. In the hadronic channel using the MLP classifier, correlated regions corresponding to $g^*\in[0.20,0.50]$ with $m_T\in[2000~\mathrm{GeV},4000~\mathrm{GeV}]$ can be excluded at the $2\sigma$ level, while the discovery reach at $5\sigma$ extends to $g^*\in[0.30,0.50]$ with $m_T$ up to approximately $3500~\mathrm{GeV}$. In the leptonic channel employing a BDT classifier, the enhanced background suppression enables exclusion and discovery sensitivity over a significantly wider mass range, with correlated regions $g^*\in[0.10,0.50]$ remaining accessible for both exclusion and discovery up to $m_T\simeq5000~\mathrm{GeV}$.

This pattern is also consistent with the closest published comparison of Ref.~\cite{Han:2025WbMuP}, where the hadronic study uses a boosted-$W$ reconstruction and reaches discovery (95\% CL exclusion) up to about $m_T\simeq3.75~\mathrm{TeV}$ ($4.5~\mathrm{TeV}$) at $\sqrt{s}=9.16~\mathrm{TeV}$ with $\mathcal{L}=100~\mathrm{fb}^{-1}$, while the present resolved-object MVA treatment gives $Z=30.20$ at $m_T=4.0~\mathrm{TeV}$ and $Z=20.28$ at $m_T=4.5~\mathrm{TeV}$ in the hadronic channel at the same collider energy, indicating numerically competitive sensitivity in the same mass region under the assumptions of this fast-simulation analysis.

Taken together, the results presented in Tables~VIII--IX and the corresponding sensitivity maps indicate a complementary two-channel pattern rather than a single universally dominant search mode.  At intermediate masses the hadronic channel benefits from the larger branching fraction and higher raw yield, while near the upper end of the scan the leptonic channel becomes more robust because of its cleaner final state and larger post-selection purity.  Any experimental interpretation of this pattern would still depend on detector-specific systematics, data-driven background constraints, and, for the hadronic channel at the highest masses, a dedicated treatment of boosted objects beyond the resolved baseline used here.

%%%%%%%%%%%%%%%%%%%%%%%%%%%%%%%%%%%%%%%%%%%%%%%%%%%%%%%%%%%%%%%%%%%%%%%%
\section{Conclusion}
%%%%%%%%%%%%%%%%%%%%%%%%%%%%%%%%%%%%%%%%%%%%%%%%%%%%%%%%%%%%%%%%%%%%%%%%

Within the fast-simulation setup adopted in this work, single production of a singlet vector-like $T$ quark at a future $\mu p$ collider remains testable in the $T\to Wb$ mode over a broad region of mass and coupling space.  The comparison with the closest existing studies shows explicitly that the collider energies and final states considered here are not new by themselves; the quantitative difference arises from the multivariate signal-extraction strategy.  At the common benchmark point $m_T=3~\mathrm{TeV}$, $\sqrt{s}=9.16~\mathrm{TeV}$, and $\mathcal{L}=3000~\mathrm{fb}^{-1}$, recalculation from the published cut-based yields gives $S/\sqrt{S+B}=18.05$ with $Z_A=1.44$ in the hadronic channel and $S/\sqrt{S+B}=3.51$ with $Z_A=0.330$ in the leptonic channel once a $20\%$ background uncertainty is included.  For the corresponding MVA working points in the present analysis, these values become $273.51$ and $42.24$ in the hadronic channel, and $113.85$ and $21.51$ in the leptonic channel, showing that the main improvement is the large increase in post-selection purity under a common significance prescription.

The full scan over $m_T=2$--$5$~TeV and the $(g^{*},m_T)$ plane shows a clear complementarity between the two final states.  At $\sqrt{s}=9.16~\mathrm{TeV}$ and $\mathcal{L}=3000~\mathrm{fb}^{-1}$, the hadronic channel with the MLP classifier excludes regions with $g^{*}\in[0.20,0.50]$ up to $m_T\approx4.0$~TeV and reaches $5\sigma$ discovery for $g^{*}\in[0.30,0.50]$ up to about $3.5$~TeV, while the leptonic channel with the best-performing classifier retains both exclusion and discovery sensitivity for $g^{*}\in[0.10,0.50]$ up to $m_T\approx5.0$~TeV.  The hadronic mode therefore benefits from the larger branching fraction at intermediate masses, whereas the leptonic mode becomes the more robust probe toward the mass frontier because of its cleaner final state and higher signal purity.

For the closest published hadronic comparison, Ref.~\cite{Han:2025WbMuP} reports at $\sqrt{s}=9.16~\mathrm{TeV}$ and $\mathcal{L}=100~\mathrm{fb}^{-1}$ a boosted-object discovery reach up to about $m_T\simeq3.75~\mathrm{TeV}$ and a 95\% CL exclusion reach up to about $m_T\simeq4.5~\mathrm{TeV}$, whereas the present resolved-object MVA study gives $Z=30.20$ at $m_T=4.0~\mathrm{TeV}$ and $Z=20.28$ at $m_T=4.5~\mathrm{TeV}$ in the hadronic channel at the same collider energy; this supports numerically competitive sensitivity through the $4.0$--$4.5~\mathrm{TeV}$ region under the present assumptions, while the comparison is not strictly one-to-one because the hadronic $W$ is reconstructed with a boosted fat-jet strategy in Ref.~\cite{Han:2025WbMuP} and with a resolved small-$R$ strategy here.

These results should be interpreted as phenomenological projections under the stated generator, detector, resolved-reconstruction, and background-uncertainty assumptions.  In particular, the hadronic reconstruction at the highest masses is a resolved-object baseline rather than a dedicated boosted-$W$ analysis, so a detector-specific treatment of boosted objects and data-driven background control could modify the absolute reach.  Subject to these limitations, the main conclusion of this work is that a systematic multivariate treatment materially strengthens the projected $\mu p$ sensitivity to $T\to Wb$ relative to the available cut-based baseline, while preserving a transparent benchmark-level comparison to the existing literature.

Finally, single-production searches at the LHC already probe part of the lower-mass region of the parameter space considered here \cite{CMS:2026vlqWb}.  Because those results are expressed in terms of a different coupling convention ($\kappa_W$) from the coupling $g^{*}$ used in this work, a quantitative mapping between the two is beyond the scope of this study, and the sensitivities reported here should be read as projections for a future $\mu p$ collider rather than as a reinterpretation of existing LHC bounds.

%%%%%%%%%%%%%%%%%%%%%%%%%%%%%%%%%%%%%%%%%%%%%%%%%%%%%%%%%%%%%%%%%%%%%%%%
\begin{table}[h!]
\centering
\caption{Benchmark scenarios for the singlet vector-like T-quark used in this analysis.  The displayed $Wb:Zt:Ht=50:25:25$ branching fractions denote the heavy-mass asymptotic singlet assumption.}
\begin{tabular}{|c|c|c|c|c|}
\hline
Benchmark & $m_T$ (GeV) & $|V_{4b}|$ & $(\kappa_W, \kappa_Z, \kappa_H)$ & Branching Ratios (\%) \\
\hline
BP1 & 2000 & 0.1 & (1,1,1) & $Wb: 50,\; Zt: 25,\; Ht: 25$ \\
BP2 & 3000 & 0.1 & (1,1,1) & $Wb: 50,\; Zt: 25,\; Ht: 25$ \\
BP3 & 4000 & 0.1 & (1,1,1) & $Wb: 50,\; Zt: 25,\; Ht: 25$ \\
BP4 & 5000 & 0.1 & (1,1,1) & $Wb: 50,\; Zt: 25,\; Ht: 25$ \\
\hline
\end{tabular}
\label{tab:benchmark}
\end{table}
%%%%%%%%%%%%%%%%%%%%%%%%%%%%%%%%%%%%%%%%%%%%%%%%%%%%%%%%%%%%%%%%%%%%%%%%
\section*{Declarations}
\subsection*{Funding}
No specific funding was received for this study.

\subsection*{Competing Interests}
The authors declare that they have no known competing financial interests or personal relationships that could have appeared to influence the work reported in this paper.

\subsection*{Data Availability Statement}
The datasets generated during and/or analyzed during the current study (Monte Carlo event samples produced with \texttt{MadGraph5\_aMC@NLO}, \texttt{Pythia8}, and \texttt{Delphes}) are available from the corresponding author on reasonable request.

\section*{AI Disclosure Statement:}
During the preparation of this manuscript, the authors utilized ChatGPT (OpenAI) to audit the bibliographic references for duplicates, verify the accuracy of internal cross-references to tables and figures, and refine the technical language and phrasing of the text. Following the use of these tools, the authors reviewed, validated, and edited the content as required and take full responsibility for the scientific results, interpretations, and the overall integrity of the publication. 

It is important to clarify that the multivariate machine learning architectures (Multilayer Perceptrons and Boosted Decision Trees) described in Sec.~\ref{sec:mva} constitute the core scientific methodology of the physics analysis and are entirely distinct from the generative AI tools used for manuscript assistance. All data simulation, processing code, and statistical analyses were developed and executed by the authors.

\end{document}